\documentclass[12pt]{article}

\usepackage{scicite}

\usepackage{graphicx}
\graphicspath{{figures/}}

\usepackage{ulem}

\usepackage{amssymb}
\usepackage{amsmath}
\usepackage{amsbsy}

\usepackage{booktabs}
\usepackage{siunitx}
\usepackage{caption}
\usepackage{color}
\usepackage[dvipsnames]{xcolor}

\DeclareSIUnit{\calorie}{cal}

\newcommand{\di}{\mathrm{d}} 

\usepackage{times}

\newenvironment{sciabstract}{%
\begin{quote} \bf}
{\end{quote}}

\title{Long-term memory and synapse-like dynamics in two-dimensional nanofluidic channels}

\author
{P. Robin$^{1,\dag}$, T. Emmerich$^{1,\dag}$, A. Ismail,$^{2,3,\dag}$,\\
 A.  Nigu\`es$^{1}$, Y. You$^{2,3}$, G.-H. Nam$^{2,3}$, A. Keerthi$^{2,4}$, \\
A. Siria$^{1}$, A.K. Geim$^{2,3}$, B. Radha$^{2,3,\ast}$, L. Bocquet$^{1,\ast}$\\
	\normalsize{$^{1}$Laboratoire de Physique de l'Ecole normale Sup\'erieure, ENS, Universit\'e PSL}, \\
	\normalsize{CNRS, Sorbonne Universit\'e, Universit\'e de Paris, 75005 Paris, France}\\
	\normalsize{$^{2}$ National Graphene Institute, The University of Manchester, Manchester, UK} \\
	\normalsize{$^{3}$ Department of Physics and Astronomy, The University of Manchester, Manchester, UK}\\
	\normalsize{$^{4}$Department of Chemistry, University of Manchester, Manchester, UK}\\
	\normalsize{$^\ast$To whom correspondence should be addressed; }\\
	\normalsize{E-mail:  lyderic.bocquet@ens.fr,radha.boya@manchester.ac.uk}\\
	\normalsize{$^\dag$ Equal contributions}
}

\date{}

\newcommand{\rev}[1]{{\color{black} #1}}

\newcommand{\Du}{\mathrm{Du}}

\usepackage[section]{placeins}

\begin{document} 




\maketitle 

\textbf{One Sentence Summary:}  
Electrolytes in 2D nanochannels develop long-term memory, allowing to implement Hebbian learning on a nanofluidic chip. 
\begin{sciabstract}
{Fine-tuned ion transport across nanoscale pores is key to many biological processes such as neurotransmission. Recent advances have enabled the confinement of water and ions to two dimensions, unveiling transport properties unreachable at larger scales and triggering hopes to reproduce
the ionic machinery of biological systems. Here we report experiments demonstrating the emergence of memory in the transport of aqueous electrolytes across (sub)nanoscale channels.
We unveiled two types of nanofluidic memristors, depending on channel material and confinement, with memory from minutes to hours. We explained how large timescales could emerge from interfacial processes like ionic self-assembly or surface adsorption. 
Such behavior allowed us to implement Hebbian learning with nanofluidic systems. This result lays the ground for biomimetic computations on aqueous electrolytic chips. }


\end{sciabstract}

Over the past decade, research in nanofluidics has shed the light on many unconventional phenoma arising in the transport of water and ions through nanometric  channels {\cite{kavokine_fluids_2021,bocquet_nanofluidics_2020,wang2020anomalies,celebi_ultimate_2014,marcotte_mechanically_2020,garaj_graphene_2010,merchant2010dna,schneider2010dna,feng_single-layer_2016,secchi_massive_2016,tunuguntla_enhanced_2017,kavokine2022fluctuation}}. 
The field has grown at a fast pace, driven by the discovery of new fundamental behaviors of aqueous transport at nanoscales, but also by their potential for a wealth of applications, from water desalination to energy harvesting \cite{bocquet_nanofluidics_2020}. Most notably, the recent development of two-dimensional (2D) channels made by van der Waals assembly of various materials (graphite, hexagonal boron nitride, MoS$_2$, etc.) has enabled the study of ionic transport at the smallest scales, with unmatched versatility in terms of geometry or surface properties \cite{radha_molecular_2016,esfandiar_size_2017,mouterde_molecular_2019,emmerich2022}. Specifics of 2D interactions offer a new asset to fine-tune the properties of electrolytes, at odds with their bulk response.   
A recent, noticable prediction is that two-dimensional ionic self-assembly should be at the root of memory effects
associated with conductance hysteresis under electrical forcing  \cite{robin_modeling_2021}, a phenomenon known as memristor effect. \rev{This effect could allow to emulate the brain's neuronal computation} using ions in water as charge carriers, but artificial systems capable of mimicking this behavior have eluded experimental inquiry in aqueous electrolytes until now.

A memristor -- short for memory resistor -- is a resistor with an internal state that is susceptible to change depending on the history of voltage seen by the system, thereby modifying its conductance \cite{chua_memristor-missing_1971,strukov_missing_2008}. As this feature makes them the analogues of biological synapses, memristors have drawn considerable attention for their potential use as building blocks of bio-inspired neuromorphic computers\cite{sebastian_memory_2020}. However, most of existing examples are based on solid-state devices (like the metallic-insulator-metallic, or MIM, architecture) and function with coupled ion and electron dynamics \cite{ge_atomristor_2018}. Although a handful of fluidic memristors were also designed {\cite{bu2019nanofluidic,sheng_transporting_2017,zhang_nanochannel-based_2019}}, they require high voltage to operate, well above the water splitting threshold ($1.23 \, \si{\volt}$ with respect to normal hydrogen electrode), use non-aqueous environments, or far exceed the nanoscale dimensions of biological systems. 
More generally, a challenge is
to replicate the mechanism found in biological systems, where the transport and accumulation of solvated ions (notably calcium) in water are used for signalization, information processing and the building of memory \cite{hille_ionic_1978,gerstner_spiking_2002}. Developing such bio-inspired memristors would notably allow to design artificial nanofluidic chips for neuromorphic computation, build an interface between artificial nanofluidics and biological systems and  explore possible gains in efficiency from using solvated ions as charge carriers.
Here we report on a series of experiments that 2D nanofluidic channels do open this avenue towards neuromorphic iontronics.

\section*{Experimental demonstration of nanofluidic memristors}

\subsection*{Pristine MoS$_2$ channels vs. activated carbon channels}
In this work, we investigated two types of 2D nanochannels, of similar geometry but different surface properties (Fig. 1A). ``Pristine'' channels were made of two atomically smooth flakes of 2D material (here MoS$_2$) separated by an array of multiple layers of graphene nanoribbons used as spacers. On the other hand, ``activated'' carbon channels consisted of two graphite flakes, in which a nanoscale trench was milled into the bottom flake using electron-beam-induced etching (EBIE) \cite{emmerich2022}. In both cases, the bottom wall of the channel was pierced and deposited on the aperture of a SiN$_x$ membrane. Further details regarding the fabrication of activated and pristine channels can be found in Refs. \cite{radha_molecular_2016} and  \cite{emmerich2022} respectively, and recalled in Supplementary Material (SM, Figs. S1, S2). 
Although similar in design, these channels differed on a few key properties. The height of pristine MoS$_2$ channels could be precisely controlled in increments of  $0.34 \, \si{nm}$, and here down to {$0.68 \,\si{nm}$} -- the channel's depth corresponding to the spacers' thickness. Conversely, the depth of activated carbon channels was controlled by EBIE with a resolution limited to a few nanometers. As recently evidenced by Emmerich et al. \cite{emmerich2022}, the latter carries a much stronger surface charge compared to pristine walls, due to the exposure of their bottom wall to the electron beam. Here, we used activated carbon channels with channel height between 4 and $13 \, \si{nm}$, and pristine MoS$_2$ channels with height between {$0.68$} and $86 \, \si{nm}$. 

\begin{figure}
	\centering
	\includegraphics[width=1\linewidth]{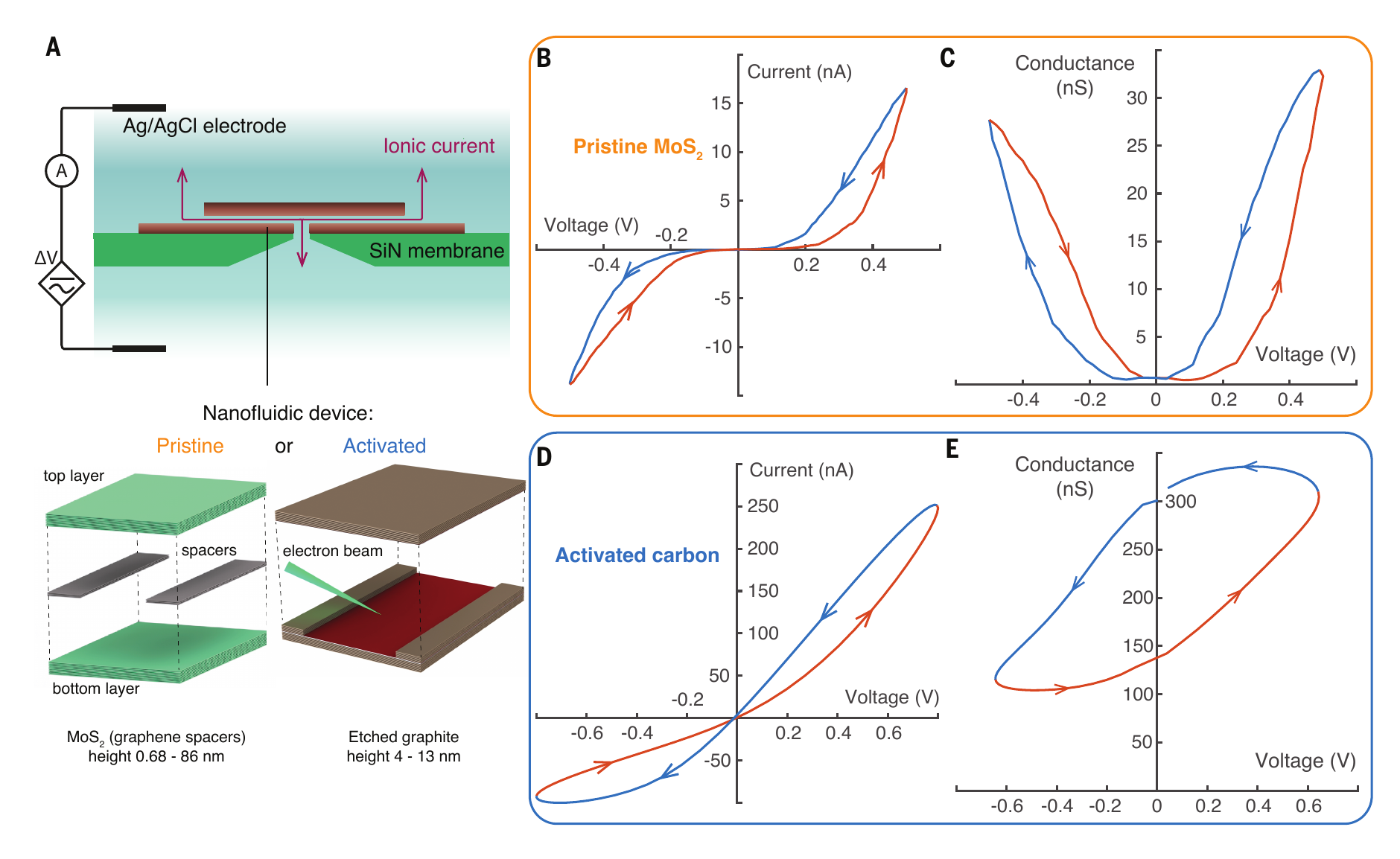}
	\caption{ \textbf{Experimental study of the memristor effect using two kinds of nanofluidic devices.} (\textbf A) Sketch of the nanofluidic cell. A nanochannel was deposited on a membrane separating two reservoirs filled with an electrolytic solution. The red arrow indicates path taken by water and ions accross the system. \rev{We used two types of nanochannels: pristine MoS$_2$ channels (bottom left pannel) and activated carbon channels (bottom right pannel).} (\textbf{B} and \textbf C) Typical example of a memristive current-voltage and conductance-voltage characteristics of a pristine MoS$_2$ channel (height $h = 1 \, \si{nm}$) under periodic voltage (triangular wave of frequency $8 \, \si{mHz}$, using $3 \, \si{M}$ KCl). The IV curve displays a loop that is pinched (but does not intersect) at the origin, and the GV curve has {a crossing point at zero voltage}. (\textbf{D} and \textbf E) Typical example of a memristive current-voltage and conductance-voltage characteristics of an activated carbon channel (height $h = 13 \, \si{nm}$) under periodic voltage (sinusoidal wave of frequency $1 \, \si{mHz}$, using $1 \, \si{mM}$ CaCl$_2$). Here, the IV curve crosses itself at the origin and the GV curve takes the form of a simple loop. For all data, pH was not modified after salt dissolution in deionized water, resulting in a pH range of $5.1-5.5$.}
\end{figure}

Once fabricated, the 2D channels were embeded in a fluidic cell connected to two reservoirs filled with electrolyte (KCl, NaCl, LiCl, CaCl$_2$, NiSO$_4$, AlCl$_3$). Salt concentrations between $1 \, \si{mM}$ and $3 \, \si{M}$ were tested. {Unless stated otherwise, the solution's pH was not modified after salt dissolution in deionized water, resulting in a pH range of $5.1-5.5$ depending on salt concentration.}
A patch-clamp amplifier (\textit{KEITHLEY} 2400 or 2600 Series) connected to Ag/AgCl electrodes allowed for ionic current measurements under imposed time-dependent voltage drop $V(t)$ of various frequencies ($0.1 - 200 \, \si{mHz} $), shapes (sinusoidal, triangular), and amplitude ($0.1 - 1 \, \si{V}$). In each case, the channel's conductance $G(t)$ was determined from current measurements from an instantaneous Ohm's law $G(t) = I(t)/ V(t)$.
Further details regarding current measurements are reported in {Supplementary Materials (SM)}. \rev{Typical examples of experimental results for both types of systems are shown in Fig.~1.}

\subsection*{Two types of memristors}

{Firstly, our central} result is that 2D nanofluidic channels did exhibit a memristive effect (Fig.~1B-E): when probed by a time-varying voltage, they displayed a non-linear current-voltage characteristics which was pinched at zero voltage, associated with a conductance hysteresis. This pinched loop {under periodic forcing} is the hallmark of memristors {\cite{pershin2011memory}}. Such behavior was found in both types of channels -- pristine and activated -- for all tested {electrolytes and at all salt concentrations}; see SM for exhaustive results. The memristive effect was found to take place at frequencies between $0.1$ and $200 \, \si{mHz}$, well below frequencies where capacitive effects can introduce hysteresis. This result corresponds to dynamical timescales from seconds to hours. {Moreover, the phenomenon was found to be robust and was observed in a wide range of parameters -- notably salt concentration, channel height and pH} {(Figs. S9-S16). All tested salts displayed the same phenomenology. In particular, we did not observe significantly different results with multivalent salts, suggesting that the materials used here (MoS$_2$ and activated carbon) are not subject to phenomena like charge reversal commonly observed with divalent ions like calcium \cite{van2006charge}. Our channels typically displayed an overall conductance much higher than what could be expected from bulk estimates, due to their high surface charge and hydrodynamic slippage \cite{emmerich2022}.}

Secondly, we could identify two types of memristors, depending on whether the current-voltage characteristics did, or did not self-cross at the origin, see Fig.~1B (pristine MoS$_2$ channel) versus {Fig.}~1D (activated carbon channel). This fundamental difference is best illustrated by looking at the curve of conductance as function of voltage: it either displayed a {twisted loop (with one crossing point, Fig.~1C) or open loop (no crossing point, Fig.~1E)}. {Following the terminology introduced by Ref. \cite{pershin2011memory}, we classified our experimental data as follows. Systems that exhibited a self-crossing IV curve (Fig.~1D and 1E) were termed bipolar memristors. Conversely, \rev{those} that instead displayed a self-crossing GV curve (Fig.~1B and 1C) were refered to as unipolar memristors.}

A key aspect of memristors is their ability to \rev{switch} between different internal conductance states. {We observed that bipolar memristors change state when the polarity of voltage was reversed, with e.g. maximum conductance at $+1\,$V and minimum conductance at $-1\,$V (Fig.~1E) -- hence the name bipolar. On the other hand, unipolar memristors generally exhibited symmetric IV and GV curves, that were however strongly non-linear when the amplitude of voltage was increased. Their conductance only weakly depended on voltage polarity, but could vary by up to two orders of magnitude between voltage $V=0$ and $V = \pm 1\,$V (Fig.~1C) -- hence the name unipolar.}
Together, these facts \rev{indicate} that the possible internal states of {unipolar} and {bipolar} memristors were fundamentally different.

In addition, thinner pristine devices (channel height $h < 10 \, \si{nm}$) could display either kind of behaviour depending on salt concentration (with unipolar memristors at 0.1 M or higher). Thicker pristine MoS$_2$ channels, on the other hand, only displayed {a weak bipolar memristor effect}; however, the memory effect was not as significant as that observed in thinner channels implying that {confinement in a 2D geometry} is essential for attaining memory effects. {Lastly, the phenomenon was found to be weakened at acidic pH in both types of systems.} All corresponding data are provided in SM \rev{(Figs. S9 to S16)}.

This comparison sheds light on a possible explanation. {Bipolar} memristors were predominant for high surface charges (as found in activated carbon channels) and low {salt} concentration -- in other words, for surface-dominated conduction. Instead, {unipolar memristors} existed for moderate surface charge (pristine MoS$_2$ channels), high {salt} concentration and strong confinement: these results corresponded to a `confinement-dominated' regime. In both cases, a prerequisite for memory effects was the system's ability to display non-linear ion transport. Accordingly, we now focus on the description of the system's various conductance states as function of applied voltage.

\subsection*{Two sources of non-linearity: collective ionic transport and ionic rectification}

The above observations suggested the existence of two distinct mechanisms behind the memristive behavior of nanofluidic channels. {Unipolar memristors} only existed in thin channels ($h < 10 \, \si{nm}$) and at high {salt} concentration ($c \geq 0.1 \, \si{M} $) and the corresponding experimental results  directly echoed the theoretical mechanisms discussed in Ref. \cite{robin_modeling_2021}.
In this picture, a non-linear response can be accounted for by the formation of tightly bound Bjerrum pairs of ions
if confinement is sufficiently strong (and the solution not too diluted), preventing conduction (Fig. 2A). The application of a sufficiently strong electric field can break these pairs or assemble them into an arc-like polyelectrolyte, allowing \rev{in both cases} electrical current to flow, a process known as the (second) Wien effect. As a result, the system's conductance $G$ is a strongly non-linear function of voltage $V$ that almost vanishes in absence of voltage, behaving as
\begin{equation}
	G(V) \propto |V|^\alpha
	\label{eqn:GWien}
\end{equation}
with a predicted exponent \rev{$\alpha > 1$}, and usually around 2 (see SM, section 3 and Ref. \cite{robin_modeling_2021} for the derivation). \rev{We can take into account the fact that not all ions may be paired up} by adding a small constant term $G_0 = G(V=0)$ into the above equation. This mechanism is independent of voltage sign, and thus does correspond {to a unipolar memristor}. It also allows the conductance to vary continuously over a large range of values, accounting for experimental observations. In theory, this process can only take place in thinner channels -- less than $2 \, \si{nm}$ in thickness -- as Bjerrum pairs only exist under strong confinement. In practice, the transition from {unipolar to bipolar} behavior was found to take place around a thickness of $10 \, \si{nm}$. A possible explanation for this robustness is that 
ion pairs could still exist in the few water layers next to the channel's walls, even in slightly larger channels: \rev{in particular, the presence of a wall tends to lower the dielectric constant of the first water layers \cite{fumagali2018anomalously}, and ions are thus expected to experience stronger electrostatic interactions near surfaces. If that is the case, then ionic pairing near solid surfaces could be relevant in other contexts and their dynamics could be probed for with similar time-varying voltage.} 

On the other hand, {bipolar} devices changed state depending on the {polarity} of applied voltage, and their memory should therefore stem from an internal asymmetry. However, {in some experimental conditions}, pristine MoS$_2$ channels did display this kind of hysteresis despite their internal surface being atomically smooth and controlled. Therefore, we attributed the source of asymmetry to entrance effects. By construction, the \rev{SiN$_x$} membrane was present on one side of the device only (Fig. 1A) and the two mouths of the channel did not have the same access resistance. Coupled with the exclusion of anions from the channel (due to its strong negative surface charge), this result is expected to result in ionic rectification (Fig. 2B, Fig. S6): when cations flow from the side with lower access resistance, ions will accumulate inside the channel as entry is `easier' than exit, resulting in a conductance increase.  If voltage is reversed, cations will flow from the side of higher resistance, the channel will instead be depleted and conductance will drop. This mechanism is analogous to that of a PN junction, and results in a diode-like current-voltage characteristic \cite{bocquet_nanofluidics_2010} with two distinct possible values of conductance, defining a rectification factor $\beta_\text{Rect}$: 
\begin{equation}
	\frac{G(V>0)}{G(V<0)} \simeq \beta_\text{Rect}
	\label{eqn:GRect}
\end{equation}
{Experimentally, we found $\beta_\text{Rect} = 1 - 5$, consistent with the above analysis in terms of entry effects (see SM, section 3)}. Because this type of non-linearity depends on voltage sign, it {corresponds to a bipolar memristor}.

{We stress that these two phenomena were not mutually exclusive: pristine MoS$_2$ channels could show signs of both mechanisms at the same time. In such cases, the IV curve displayed two crossing points (rather than none or a single one); further analysis and experimental data can be found in SM (section 3.4 and Fig.~S17).}

Although any system presenting {a strong enough non-linearity associated with various internal conductance states} could in theory display a memristive behavior, it can only do so on a frequency range fixed by the time  {required} to switch between the conducting and the insulating states. However, such timescales are normally too short to be accessible in {nanofluidic} systems, and this phenomenon requires peculiar transport processes to be observed.

\begin{figure}
	\centering
	\includegraphics[width=1\linewidth]{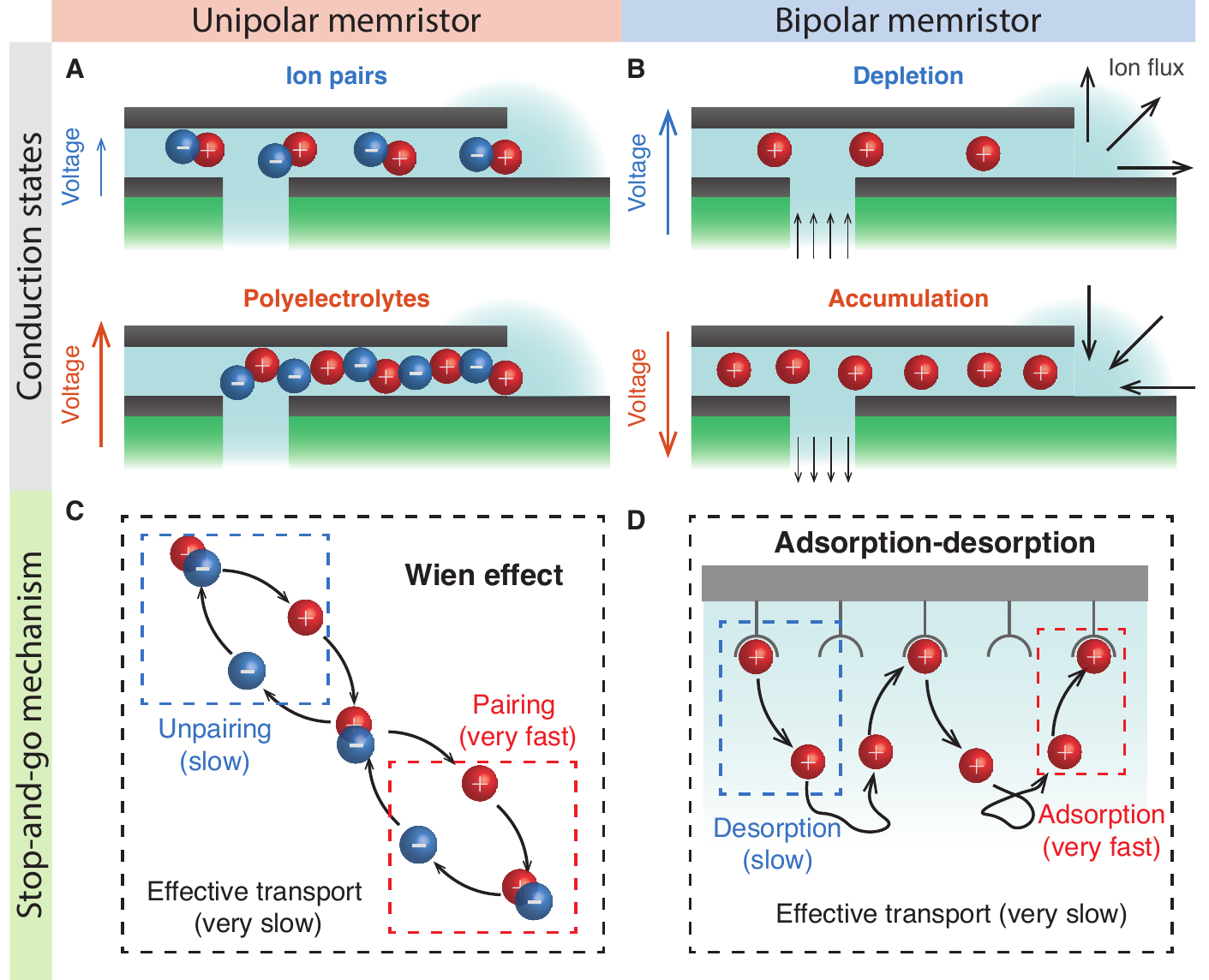}
	\caption{ \textbf{Theory of nanofluidic memristors.} (\textbf{A} and \textbf B) \rev{Description of the different conductance states explaining the memristor effect.} (\textbf{A}) Wien effect as a source of {the unipolar memristor effect}. Under strong confinement, ions assemble into non-conducting Bjerrum pairs. Conduction can then occur through the breaking of pairs (Wien effect) or their clustering into conducting arcs (polyelectrolytic Wien effect), under the action of a strong electric field (regardless of sign). (\textbf{B}) Entry effects as a source of {the bipolar memristor effect}. The two mouths of the channel are asymmetric, resulting in ionic rectification depending on the side from which charges enter the system. If they enter from the side of low resistance, ions accumulate and conductance rises. Otherwise, the channel is depleted and conductance is lowered. {As the channel's walls bear a strong negative charge, only positive ions are represented here.} (\textbf{C} and \textbf D) {Effective `stop-and-go' transport} and long-term memory. In both mechanisms (Wien effect or geometrical asymmetry), the system's conductance state can be retained over large timescales if transport is governed by a stop-and-go motion, induced by repeated pairing/unpairing or adsorption/desorption events\rev{, respectively}.}
\end{figure}

\subsection*{Stop-and-go transport as a source of long-term memory}

For both {types of} memristors, memory timescales were found to reach extremely large values, in the range of minutes to hours. 
Such long-term memory could be accounted for by taking ion pairing or surface adsorption into account in the dynamics of confined ions.
In the theoretical framework of Ref. \cite{robin_modeling_2021}, the electrolyte is indeed predicted to retain its conductance state through the formation of ion clusters, which was estimated to typically take a few milliseconds. More generally, one expects a nanofluidic channel to retain a conductance state (defined by the number of charge carriers present inside the channel) over a typical diffusion timescale, roughly $L^2/D$, with $L$ the channel length and $D$ a typical ionic diffusion coefficient in the channel. For channels of length around $10 \, \si{\micro \meter}$, this result would yield a maximum memory time of $0.1 \, \si{s}$, still orders of magnitude lower than experimental values. However, this picture changes if interfacial processes, rather than diffusion, govern ion transport. Consider a particle diffusing through a channel with chemically-active walls, such that it may adsorb on the surface (Fig. 2D). If the adsorption rate is much larger than the diffusion rate across the channel, then the particle will spend most of its time bound to the surface. As a result, the time it needs to escape the pore is the sum of the durations of all adsorption events. Let us define $\tau_\text{diff}$, the time needed to escape through diffusion alone and $\tau_d$ the time a particle bound to the surface takes to desorb. Then, if the particle is adsorbed every $\tau_a \ll \tau_\text{diff}$, there will be $\tau_\text{diff}/\tau_a$ such events along the particle's trajectory as it {crosses} the channel. As a result, the {residence} time $\tau_m$ of the particle inside the pore reads:
\begin{equation}
	\tau_m = \frac{\tau_\text{diff}}{\tau_a} \tau_d = \frac{\tau_d}{\tau_a} \tau_\text{diff} \gg \tau_\text{diff}
	\label{taum}
\end{equation} 
In other words, the memory time of the system is the diffusion timescale times a ratio $\tau_d/\tau_a$ measuring the strength of surface effects. At chemical equilibrium, this condition corresponds to the ratio of particle numbers on the surface and in the bulk of the channel, as quantified by \rev{the dimensionless} Dukhin number $\Du = \Sigma/ehc$, which compares the surface charge $\Sigma$ to the charge density in the bulk of the electrolyte, $ec$. Putting numbers, activated carbon channels typically have $\Du \sim 10^2-10^3$, showing that surface effects strongly dominate the bulk. Eq.(\ref{taum}) then predicts a  memory time in the range $\tau_m\sim \Du \times \tau_\text{diff}\sim$ 100s. This estimation is in agreement with experimental values, which were found to be in the range $\tau_m \sim 50-400 \, \si{s}$ (Figs. S12 and S14){, and is consistent with previous reports of extremely slow diffusion of ions near a chemically active surface \cite{comtet2020direct}}. {Our prediction was found to generally underestimate experimental values: this can be attributed to the fact that our model assumes independent successive adsorption events, while in reality they tend to be correlated over long timescales \cite{gravelle2019adsorption}.} {Moreover, this surface-driven mechanism could explain the disappearance of the phenomenon at low pH (Fig.~S16), which is known to greatly influence the channels' surface charge \cite{emmerich2022}. The observed dependence of the memristor effect with the electrolyte could likewise result from difference in chemical affinity between the various species of ions considered and surface defects (Figs. S9 and S14) -- however, this depence is hard to analyze and would require further knowledge of the chemical nature of adsorption sites.}

{We note that this slow `stop-and-go' motion of ions near the channel's surface is not incompatible with the high conductance of some (notably activated) channels: although surface processes such as adsorption can slow down conductance changes, they do not modify the overall large number of ions present in the channel due to its strong surface charge.} \rev{We recall how to link conductance to surface charge in SM (section 3.2). Similarly, we observe that the slow down of the dynamics on the timescales of minutes emerges from microscopic processes (adsorbing events) with molecular timescales (say $1 \, \si{\micro \second}$ at best). This is, however, reminiscent of previous studies that showed how chemical or physical surface processes, involving notably the Stern layer, can result in hour-long phenomena when coupled to a water flow \cite{ober_liquid_2021,werkhoven_flow_2018}.}

A similar argument can be formulated for {unipolar memristors}. This time, the conduction state of the system is encoded in the number of ions which are not part of tightly bound pair (and can therefore contribute to current) -- according to the Wien effect. Similarly to surface adsorption, one expects that successive pairing-unpairing events will create 
a stop-and-go motion of ions through the system (Fig. 2C). The memory time is then again found to be given by  diffusion times a ratio of pairing and unpairing times, potentially reaching minute- or even hour-long timescales.

\begin{figure}
	\centering
	\includegraphics[width=1\linewidth]{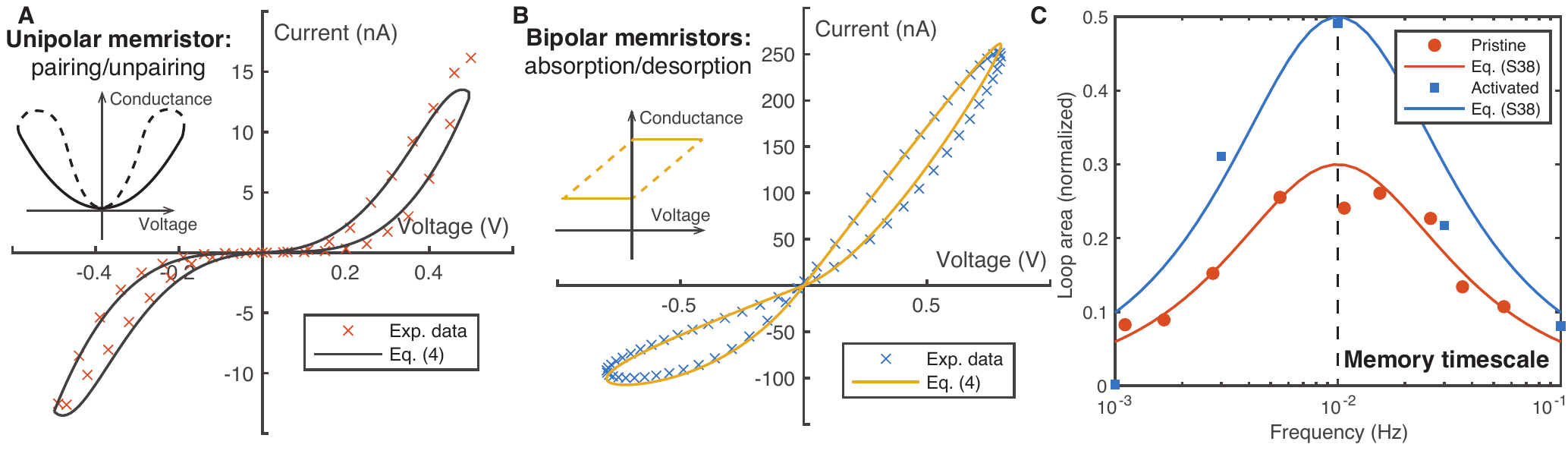}
	\caption{ \textbf{Comparison of {theoretical models} and experimental results.} (\textbf{A} and \textbf B) Fit of experimental {IV curves} using the minimal model of the nanofluidic memristor. For {unipolar memristors} (A), the quasistatic conductance was taken to be a power law of applied voltage (here with exponent $\alpha=2$), and a sign-dependent constant {in the bipolar case} (B), see insets. The experimental curves were then fitted using the delay time $\tau$ as single free parameter, see Eq. \eqref{eqn:ConvolG}. Datasets correspond to devices presented in Fig. 1B-E: $3 \, \si{M}$ KCl in pristine MoS$_2$ channel (height $1 \, \si{nm}$) or $1 \, \si{mM}$ CaCl$_2$ in activated carbon channel (height $13 \, \si{nm}$). (\textbf{C}) Normalized area of the IV loop as function of voltage frequency. Data correspond to $100 \, \si{mM}$ CaCl$_2$ in $4 \, \si{nm}$ activated carbon channel and $1 \, \si{M}$ KCl in $0.68 \, \si{nm}$ pristine MoS$_2$ channel. The memory timescale $\tau_m$ could be extracted from experimental data by looking for the frequency where the loop was the largest. The curve of loop area as function of frequency was well described by that of the minimal model{, see Eq. (S38) of SM} (solid lines). See Fig. S8 for the normalization process.}
\end{figure}

Building on this qualitative picture, one may propose a minimal model, accounting for the memristor effect over minute-long times for {both memory types}, as detailed in SM (Fig. S7). We found that the system's conductance at time $t$ was given by the convolution of its quasistatic conductance, as given by Eqs. \eqref{eqn:GWien} and \eqref{eqn:GRect} depending on memristor type, with an exponential memory kernel:
\begin{equation}
	G(t) = \int_0^{\infty} G_\text{qs}[V(t-s)] \frac{e^{-s/\tau}}{\tau} \, \di s 
	\label{eqn:ConvolG}
\end{equation}
where $G_\text{qs}$ is quasistatic (non-linear) conductance and $\tau$ a timescale of the order of the memory time $\tau_m$. The resulting prediction was in good agreement with experimental data (Fig. 3A-B). According to this simple model, measuring the loop in the IV curve allowed to characterize the memristive effect (Fig.~3C). The curve of area as function of voltage frequency exhibited a maximum when the frequency matched the intrinsic memory time $\tau_m$, akin to a resonance. The comparison to the prediction of the model showed again a good agreement and provided a direct measurement of $\tau_m$ (Fig. 3C).

\section*{Hebbian learning with nanofluidic memristors}

\subsection*{Reversible {modification of a nanochannel's conductance}}

This qualitative and quantitative rationalization of the ionic memristor effects paves the way for the implementation of learning algorithms using our nanofluidic devices. As a proof of concept, we now show that they could be used to emulate some basic functionalities found in biological synapses. Because their memory was not lost when voltage was reset to zero, we only focused on {bipolar memristors,} as exhibited here with activated carbon channels. We first confirmed that their conductance could be increased or lowered through {successive} voltage sweeps of a given {polarity} {(Fig.~4A)}. Following a positive spike, the conductance was abruptly increased for a short period ($\sim 1 \, \si{min}$), before relaxing to a long-term value above  its initial state {(Fig.~4B)}. This result shows that our device displayed both short- and long-term memory, similar to biological synapses {\cite{bao_involvement_1997}}.

\begin{figure}
	\centering
	\includegraphics[width=1\linewidth]{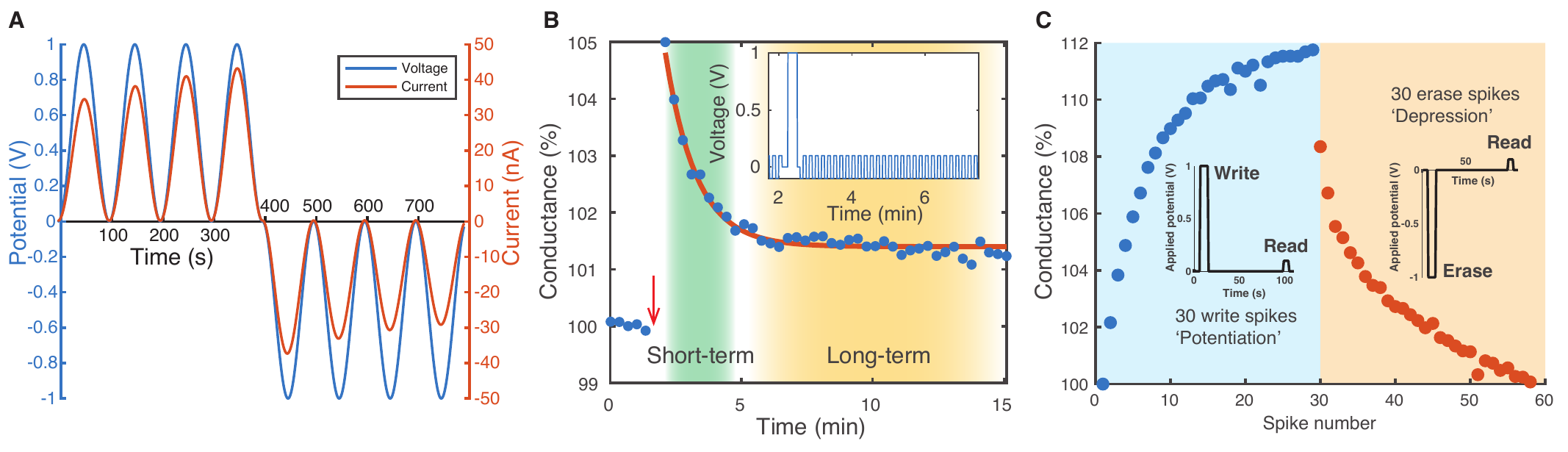}
	\caption{ \textbf{Programing a nanochannel through reversible conductance strenghening.} (\textbf A) Evolution of the ionic current (orange) under voltage pulses of constant {polarity} (blue). Positive (resp. negative) pulses result in a increase (resp. decrease) of conductance. (\textbf{B}) Conductance change following a positive voltage pulse, exhibiting both short- ($< 2 \,$min) and long-term ($> 2 \,$min) memory. The conductance was read by applying a weak square voltage wave that had no sensible impact on the state of the system, and modified through a strong voltage spike. Blue points are experimental data. The red solid line is a guide for the eye. Inset: applied voltage as function of time. The red arrow indicates the beginning of the voltage spike. (\textbf{C}) Long-term modification of a nanochannel's conductance. 30 write spikes (+1 V, 10 s) were applied, followed by 30 erase spikes (-1 V, 10 s) which brought back the system to its initial state. Between each spike, the conductance was let to stabilize during two minutes and was then measured with a read pulse (0.1 V, 5 s), see Fig. S3. All data correspond to activated carbon channels with height $h = 5 \, \si{nm}$ filled with $1 \, \si{mM}$ CaCl$_2$.}
\end{figure}

These neural connections act as resistors whose conductance can be tuned during learning processes, with reversible modifications both on short (milliseconds to minutes) and long (minutes to hours or more) timescales \cite{zucker2002short,bear2020neuroscience}. The latter, known as long-term potentiation (or depression, when the conductance is lowered) enables the storage of information through the synapse's conductance state as a form of in-memory coding. Although the exact biological mechanisms are still debated, the transport and accumulation of calcium ions at specific places play a key role  \cite{gerstner_spiking_2002, bliss_synaptic_1993}. Taking inspiration from these features, we designed a protocol to implement in-memory computations with our nanofluidic channel {(Fig.~4C)}. We incremented 
a nanochannel's conductance by applying a `write' voltage spike ($+1 \, \si{V}$ during $10 \,$seconds). It could then be accessed to via a `read' pulse ($+0.1 \, \si{V}$ during $5 \,$seconds), which did not perturb sensibly its value. It could also be reset to its original value with an `erase' spike ($-1 \, \si{V}$ during $10 \,$seconds). This setup allowed for a versatile and reversible modification and access to the stored value for computational applications. As a proof of concept, we show in Fig. 4C that the modification process was indeed fully reversible and allowed to store an analog variable over long timescales, by applying a series of 60 write and erase spikes. We thus demonstrated that nanoscale channels could be `programmed' through the tuning of their conductance, enabling the implementation of in-memory operations with ion-based nanofluidic systems.

\subsection*{Hebbian learning algorithm}

Building on the similarities between our nanofluidic system and synapses, we now implemented a basic form of Hebbian learning. In biological neuron networks, this process consists in the modification of synaptic weights depending on the relative activation timings of two neurons connected by a given synapse (Fig. 5A). If the presynaptic neuron repeatedly emits an action potential {shortly before} the activation of the postsynaptic neuron, the synapse is strengthened {(meaning its conductance is increased)}, as this result suggests a form of causality between the two activation events. Conversely, the synapse is weakened {(i.e., its conductance is decreased)} if the firing order is reversed, which would point at some anticausality relation. Importantly, these modifications occur even if the presynaptic neuron only causes a mild response (that is, too weak to initiate an action potential by itself) of the postsynaptic one. Altogether, this process implements a form of principal component analysis of the inputs received by the network \cite{gerstner2014neuronal}, and is believed to play a major role in learning.

To mimick this mechanism, we designed the experiment presented in Fig. 5B. {A computer generated a voltage time series that emulated the behaviour of two neurons. This time series was then applied on a nanofluidic channel. The activation of the presynaptic neuron A was modeled by a weak positive voltage pulse. Whenever it activated, a flip-flop mechanism was triggered, connecting the channel to a generator $E_-$ that applied negative voltage spikes.} This behaviour lasted until the {postsynaptic} neuron B activated and the system was branched on another generator $E_+$ applying positive spikes instead. The opposite chain of events occured if neuron B activated first: {in that case, the channel first received positive spikes from $E_+$, followed by negative spikes from $E_-$ once neuron A activated.} In both cases, the flip-flop reseted if a given total amount of time passed since its activation, allowing the process to start over. Further details regarding the implementation are provided in SM (Fig. S4).

\begin{figure}
	\centering
	\includegraphics[width=1\linewidth]{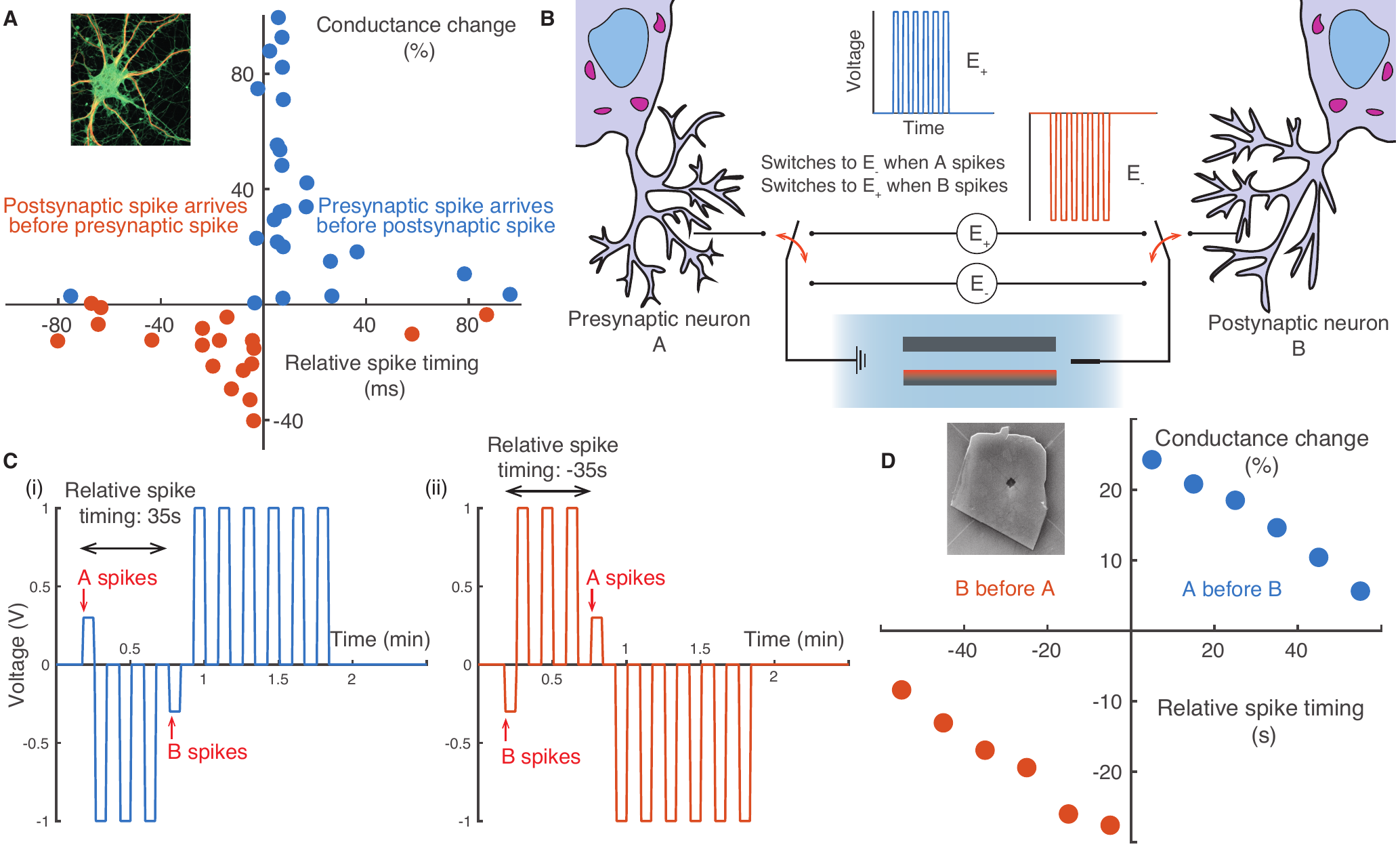}
	\caption{ \textbf{Implementation of Hebb's law using \rev{activated carbon channels.}} (\textbf{A}) Hebb's law in biological synapses: a synapse's conductance was increaseed (resp. decreased) when its presynaptic neuron fired just before (resp. after) the postsynaptic one, adapted from Ref. \cite{bi1998synaptic}. This process implemented a form of causality detection, known as spike-timing-dependent plasticity (STDP). Inset: rat hippocampal neuron (\copyright ZEISS Microscopy from Germany, CC BY 2.0). (\textbf{B}) Hebb's law with nanofluidic memristors: voltage spikes were applied to a nanochannel, mimicking the activation of two neurons A and B. After each spike from the presynaptic (resp. postsynaptic) neuron, a series of erase (resp. write) spikes was applied. (\textbf{C}) Example of voltage spikes series depending on wether the presynaptic (i) or postsynaptic (ii) neuron activated first. (\textbf{D}) Conductance change after 8 successive activations of the two neurons, in percentage of the initial conductance and as function of the relative activation timing of the neurons. {Relative spike timing is measured from the onset of the first spike to the onset of the second}. Inset: SEM image of an activated \rev{carbon }nanochannel. Data correspond to an activated \rev{carbon }channel with height $h = 5 \, \si{nm}$ filled with $1 \, \si{mM}$ CaCl$_2$. See also Fig. S5 for additional data.}
\end{figure}
If neuron A activated just before neuron B, then the system received a few negative spikes, followed by many positive spikes (Fig. 5C, left panel). Its conductance was thus globally increased. When the firing order was reversed, conversely, the system received more negative than positive spikes (Fig. 5C, right panel), and its conductance was therefore lowered.

We implemented this protocol in the experiments as follows: we first measured the system's conductance, and ran the program which consisted in 8 successive activations of neurons A and B. {Their relative spike timing -- measured from the onset of the first spike to be triggered to the onset of the second one -- was used as a tuneable parameter.} Then, we measured {the resulting} change in the conductance. The result is shown on Fig. 5D: when the presynaptic spike was followed (within a 40 seconds window) by a postsynaptic spike, {conductance was increased -- resulting in a strengthened connection between the two neurons}. Otherwise, if the delay was too great or if the order was reversed, the connection was left unchanged or weakened, respectively. This phenomenology {echoes} the one observed in biological synapses, {where the transient accumulation of certain ionic species triggers various mechanisms that ultimately result in the strengthening of neuron connections \cite{bi1998synaptic, bliss_synaptic_1993}; \rev{here, the accumulation of ions inside the nanochannel directly causes a conductance increase.}}

{In conclusion, }two-dimensional nanochannels exhibited long-term memory, in the form of a memristor effect that could have two different physical origins - strong correlations between ions or entrance effects. In both cases, memory was retained over long timescales through interfacial processes that slowed down advection-diffusion across the channel. We fully characterized experimentally and theoretically both of these mechanisms. {In particular, we showed they may be harnessed for `iontronics' applications in a variety of contexts, as the memory effect was observed in all tested experimental conditions (salt concentration, electrolyte, pH)}. These systems reproduced the tuneability of synapses, through an accumulation (or depletion) of ions, and could implement basic learning algorithms such as Hebb's rule within a simple nanofluidic architecture. More generally, our work illustrates how confinement-induced effects could be harnessed to build ionic machines inspired by biological systems. This work paves the way for the development of more complex iontronic devices on nanofluidic chips with advanced circuitry. The use of water and ions in the nanofluidic memristors, which is shared by biological systems, furthermore suggests the possibility to interface artificial with biological devices. 

\clearpage

\bibliographystyle{./Science}


\section*{Acknowledgments}

\noindent\textbf{Funding:} L.B. acknowledges funding from the EU H2020 Framework Programme/ERC Advanced Grant agreement number 785911-Shadoks and ANR project Neptune. L.B. and A.S. acknowledge support from the Horizon 2020 program through Grant No. 899528- FET-OPEN-ITS-THIN. A.K. acknowledges Ramsay Memorial Fellowship and also funding from Royal Society research grant RGS/R2/202036. B.R. acknowledges the Royal Society fellowship and funding from the EU H2020 Framework Programme/ERC Starting Grant number 852674 AngstroCAP. This work has received the support of Institut Pierre-Gilles de Gennes.

\noindent\textbf{Authors contributions:} 
{
L.B., R.B. conceived the project, designed the experiments and supervised the work, with inputs from P.R., A.K.G. and A.S.; T.E. and A.I. performed the measurements on the activated and pristine channels, respectively; A.K., Y.Y., G.-H.N., fabricated the pristine MoS$_2$ channels, T.E.,  fabricated the activated carbon channels, with inputs from A.N. and A.S.; P.R. designed the experimental protocols for neuronal mimics and developed the theoretical modelling; T.E., A.I. and P.R. analyzed the experimental data with inputs from L.B., A.S., R.B., A.K.G.. The manuscript was written by P.R., L.B., R.B. with inputs from T.E. and A.I. All authors contributed to the review and editing of the manuscript.
}

\noindent\textbf{Competing interests:} None declared.

\noindent\textbf{Data and materials availability:} {All experimental data reported here are archived on Zenodo~\cite{robin_paul_2022_7085645}. All other data needed to evaluate the conclusions in the paper are present in the paper or the Supplementary Materials.}


\section*{Supplementary materials}
Supplementary Text\\
Figs. S1 to S17\\
References \cite{bjerrum_untersuchungen_1926,prodromakis2012two,onsager_deviations_1934,siwy2002rectification,picallo2013nanofluidic,karnik_rectification_2007,poggioli2019beyond}
\end{document}



\baselineskip22pt


\maketitle 

\noindent
Supplementary Text\\
Figs. S1 to S17\\
References \cite{bjerrum_untersuchungen_1926,prodromakis2012two,onsager_deviations_1934,siwy2002rectification,picallo2013nanofluidic,karnik_rectification_2007,poggioli2019beyond}

\clearpage

\section*{Supplementary Materials}

\tableofcontents{}

\clearpage

\section{Materials and methods}

\subsection{Nanofabrication of pristine MoS$_2$ channels}
We fabricated the pristine  MoS$_2$ channels via van der Waals assembly following the protocol reported in our previous work \cite{radha_molecular_2016}. Briefly, the process has two major parts: I) the preparation of the top-spacer layers, II) the assembly of the resulting top-spacer layers with the bottom layer to form tri-crystal (top-spacer-bottom) stack. Graphene and MoS$_2$ flakes were prepared by the mechanical exfoliation of their bulk layered forms, Graphenium graphite and natural MoS$_2$ crystals (purchased from \textit{Manchester Nanomaterials}). On a Si/SiO$_2$ substrate, an exfoliated graphene flake with a specific thickness between 0.68 and 86 nm, was searched for using an optical microscope, and the flake thickness was confirmed by atomic force microscopy (AFM) (Fig.~S1A). Parallel strips (width of $\sim$120 nm, spacing $\sim$150 nm) were made from this graphene layer using e-beam lithography (EBL) and dry etching using oxygen plasma (Fig.~S1B). The spacing between the graphene strips would become the channel width $w$ in the final device, while the thickness of the graphene would become the channel height $h$. Fig. S1B (bottom panel) shows the AFM height profile of a three-layer thin graphene spacer, with channel width of $\sim$150 nm and a channel height of $\sim1.2 \pm 0.1$ nm. Following this, a MoS$_2$ crystal (thickness of 150 to 200 nm) was transferred on top of the graphene spacer (Fig.~S1C). This MoS$_2$ layer served as the top wall of the channel.
The nanofluidic chip fabrication process began by drilling a microhole ($\sim3 \, \si{\micro m} \times 50 \, \si{\micro m}$) in a SiN$_x$ membrane (thickness of 500 nm) on a silicon wafer, using photolithography and dry etching (mixture of SF$_6$ and CHF$_3$ gases). A thin MoS$_2$ layer (thickness of 20 to 40 nm) was then transferred on top of the microhole on SiNx membrane to act as the bottom layer of the channel, using polymethylmethacrylate (PMMA) based wet transfer method (Fig. S1D). Then, the bottom layer was etched from the back of the SiN$_x$ membrane via dry etching with CHF$_3$ and O$_2$ gases, to protrude the microhole onto MoS$_2$. Following this, the previously prepared stack of top MoS$_2$-graphene spacer was wet-transferred onto the bottom MoS$_2$ layer on SiN$_x$ (Fig. S1E). During this transfer, the spacer was oriented in such a way that the channels are perpendicular to the rectangular microhole. Then, the graphene spacer was dry etched (O$_2$ plasma) from the back of SiN$_x$ membrane to further open the microhole into the channels. At this stage, the resulting channels had variable lengths determined by the shape of the top MoS$_2$ flake, and any thin edges of the top layer could lead to its sagging into the channels thus blocking the channel entries. To address this, a Cr/Au strip (thickness 5 nm/70 nm, respectively) was deposited on the tri-crystal stack after photolithography to open the channels (Fig. S1E). The gold layer aided the device stability by minimizing the lifting of the MoS$_2$ top layer during measurements at high voltages. Moreover, the gold strip served as a mask to create uniform and desired channel length for all channels across the device. Regions of the tri-crystal stack not masked by the Au strip were etched away, hence only the channel region underneath the gold strip remained (Fig. S1H). Throughout the device fabrication process, after each flake transfer, the device was placed in a furnace under H$_2$:Ar (1:10) gas for annealing (300$\si{\degreeCelsius}$ for 3 hours and 400$\si{\degreeCelsius}$ for 4 hours) to clean the polymer contamination. The optical images of the final channel devices on SiN$_x$ are shown in Fig. S1I, both in reflection mode and transmission mode, with the channel length $L$ (from the microhole to the end of the Au strip) indicated.

\subsection{Nanofabrication of activated \rev{carbon} channels}

We briefly recall here the nanofabrication process of activated carbon nanochannels. A more detailed description can be found in Ref. \cite{emmerich2022}. Bidimensional graphite crystals were obtained from commercially available graphite (\textit{GRAPHENIUM}) by mechanical exfoliation on a Si/SiO$_2$ substrate using cleanroom tape. A first graphite flake (`bottom layer') was pierced using electron beam induced direct etching (EBIE) in water vapor inside a scanning electron microscope, and then several trenches were dug from the hole using the same technique (Fig. S2, step 1). A second graphite crystal (`top layer') of roughly 50 nm in thickness was deposited above the hole and covered partially the trenches, closing them to form channels (Fig. S2, step 2). This first transfer was realized using the dry-transfer techniques with a droplet-shaped polydimethylsiloxane (PDMS) stamp spin-coated with polypropylene carbonate (PPC). Finally, this bi-layer heterostructure was transferred above a Si/SiN window with a circular aperture in its middle, by making sure that the hole in the bottom layer landed above the aperture (Fig. S2, step 3). This second transfer was realized by wet transfer using a polymethylmethacrylate (PMMA) sacrificial layer. 

In typical cases, activated \rev{carbon} channels are $5-10 \, \si{\micro m}$ long, $100 \, \si{nm}$ wide, with a height ranging from 5 to $15 \, \si{nm}$.

\subsection{Current measurements}

Devices were placed into nanofluidic measurement cells separating two reservoirs filled with electrolyte solutions of various salt concentrations and ionic species (KCl, CaCl$_2$, AlCl$_3$, NiSO$_4$, NaCl, LiCl). We used Ag/AgCl electrodes to apply a potential drop across the channels and measure the resulting ionic current. Our electrodes were connected to \textit{KEITHLEY} amplifiers (models 2636B and 2401). We used AC voltage of various frequencies ($0.1$ to $200 \, \si{mHz}$) with a sampling rate of $2 \, \si{Hz}$. {Due to experimental technicalities, triangular waveforms were used to probe pristine MoS$_2$ channels and sinusoidal for activated carbon channels; it was later checked, however, that our nanochannels react identically to both types of waveforms. 
	
	Even in the thinnest channels, the measured current was found to be several orders of magnitude higher than what was measured on a control system obtained by following the ``pristine'' protocol, but without spacers to create the channels. This shows that no leakage through the channel's walls is not possible.} 

\clearpage

\section{Bio-inspired algorithms}
In this section, we detail how we performed basic neuromorphic operations with activated \rev{carbon} channels. In all cases, input voltage was generated via \textit{MATLAB} and exported as a text file, and then applied on the system by a \textit{LABVIEW} program with a sampling rate of $2 \, \si{Hz}$.

\subsection{Long-term potentiation}

Reversible long-term modification of a nanochannel's conductance was achieved by applying positive ``write'' pulses ($+1 \, \si{V}$ during $10 \, \si{s}$) or negative ``erase'' pulses ($-1 \, \si{V}$ during $10 \, \si{s}$). After each pulse, the conductance relaxation was tracked by applying 10 weak ``read'' pulses ($\Delta V_\text{read}=+0.1 \, \si{V}$ during $5 \, \si{s}$) separated by $5 \,\si{s}$ (see Fig. S2A-B), and computing for each of these pulses the conductance from:
\begin{equation}
	G = \frac{I}{\Delta V_\text{read}}
\end{equation}
The conductance was found to stabilize after roughly two minutes of relaxation.

We then checked that these modifications were incremental and reversible by applying 30 write pulses followed by 30 erase pulses, see Fig. 5C from main text. After each pulse, the system was let to relax for two minutes to let the conductance stabilize.

\subsection{Hebbian learing}

The algorithm used to implement Hebbian learning is detailed in main text but we recall it here for the sake of clarity.  {A computer generated a voltage time series that emulated the behaviour of two neurons. This time series was then applied on a nanofluidic channel. The activation of the presynaptic neuron A was modeled by a weak positive voltage pulse. Whenever it activated, a flip-flop mechanism was triggered, connecting the channel to a generator $E_-$ that applied negative voltage spikes.} This behaviour lasted until the {postsynaptic} neuron B activated and the system was branched on another generator $E_+$ applying positive spikes instead. The opposite chain of events occured if neuron B activated first: {in that case, the channel first received positive spikes from $E_+$, followed by negative spikes from $E_-$ once neuron A activated.} In both cases, the flip-flop reseted if a given total amount of time $t_0$ passed since its activation, allowing the process to start over.

To mimick Hebbian learning, we assumed that neuron B always activated with a delay $\Delta t$ compared to neuron A. Note that $\Delta t$ could be negative, \rev{if} B actually fired before A. Then, for a given value of $\Delta t$, we applied the above procedure 8 times, and measured the conductance change of the channel at the end (see Fig. S4). 

If A activates just before B, the system will be subject to a few negative spikes followed by many positive ones, increasing its overall conductance (see Fig. S4D). Likewise, If A activates just after B, the system receives a few positive spikes and many negative spikes, and its conductance is lowered. On the other hand, if $|\Delta t|$ is comparable to $t_0$, the channel will receive an almost equal amount of positive and negative spikes, leaving its conductance unchanged. 

Additional data measured on different systems than the ones presented in Fig. 5-6 from main text are displayed in Fig. S5, demonstrating the robustness of the observed phenomena.

In all cases, we prepared the channel in an intermediate conductance state so that saturation to the state of maximum or minimum conductance is not a problem during the learning process.

\section{Theory of the nanofluidic memristor}

In this section, we detail an analytical model of the nanofluidic memristor, highlighting two distinct mechanisms. The most illustrative difference between the two is the shape of the IV and GV curves under a periodic excitation. {In some cases, the GV curve displayed a self-crossing point (unipolar memristor, proposed mechanism based on Wien effect, section \ref{ssec:Wien}); in others, it had the shape of an open loop and the IV curve had a self-crossing point instead (bipolar memristor, proposed mechanism based on ionic rectification, section \ref{ssec:Rect}).}

If considered in a vacuum, both these effects only yield a memory time on the timescales of milliseconds at best. In section 2.3, we will therefore provide a minimal model explaining the emergence of long-term memory from the coupling of surface processes to bulk transport between two reservoirs.

We start by recalling that a memristor is a resistor with a hysteretic conductance. In terms of elementary electronics, it is decribed by a set of two equations:
\begin{align}
	I &= G\left[n(t)\right] \Delta V (t),\\
	\dot n &= f(n, \Delta V(t)) \label{eqn:Ndyn}
\end{align}
where $I$ is the electrical current flowing through the device under a time-varying voltage drop $\Delta V(t)$, $G$ is the conductance that depends on an internal parameter $n$, which can be seen as the system's memory. It evolves according to a dynamical equation \eqref{eqn:Ndyn}, where the dot represents the time derivative. 

The goal of this section is to detail, for the two mechanisms, what $n$ physically represents and to derive its evolution equation from the underlying physics. In other words, to explain the memristor effect observed in confined electrolytes, we need to first show that they possess several internal conductance states, and then that they are able to retain such states over long periods.
\subsection{Unipolar memristors -- Wien effect under 2D confinement}
\label{ssec:Wien}
In this section, we discuss equation $(1)$ from main text, which describes unipolar memristors (with a self-crossing GV curve). The full derivation can be found in Ref. \cite{robin_modeling_2021}: herein, we only state and discuss the various results for the sake of clarity. In this mechanism, $n$ represents the proportion of ions which are able to move under an electric field. All other ions form neutral (and therefore non-conducting) ion pairs, also refered to as Bjerrum pairs \cite{bjerrum_untersuchungen_1926}. If the field is strong enough, it will tear some pairs apart, increasing $n$ and the global conductance, in a process known as the (second) Wien effect. If the field is turned off, pairs will eventually form again. However, this pairing and unpairing process takes some time, allowing for a memristor effect, akin to an electric arc in a discharge tube: an external voltage is required to ionize the gas and make it conduct current, but the gas will stay conducting for a short time after the voltage is removed \cite{prodromakis2012two}.

The second Wien effect is a well-known phenomenon observed in weak electrolytes. It was extensively studied by Onsager for bulk electrolytes \cite{onsager_deviations_1934}, resulting in an approximate law for the conductance $G$ under an external field $E$:
\begin{equation}
	G(E) \simeq G(0)\left[1 + \frac{\beta e^2 }{4 \pi \epsilon \ell_E} + \dots\right] \simeq G(0) \left[1 + \frac{\ell_B}{\ell_E} + \dots\right]
\end{equation}
with $\beta = 1/k_B T$, $\ell_E = k_BT/eE$ a lengthscale defined by the external field and $\ell_B$ the Bjerrum length. Here, $G(0)$ corresponds to the conductance of ions that are already free at thermal equilibrium (i.e. with no external field). This results in a sligthly non-linear IV curve:
\begin{equation}
	I = G(E) \Delta V \simeq G(0) \Delta V\left[1 + \alpha \Delta V\right]
\end{equation}
$\Delta V$ being the voltage drop associated with the field $E$. Without detailing the (mathematically involved) exact derivation by Onsager, the critical point is that ion pairs form according to a chemical equilibrium given by:
\begin{equation}
	\dot n_p = \frac{1}{\tau_a} n_f^2 - \frac{1}{\tau_d}n_p
\end{equation}
with $n_p$ the proportion of ion pairs, $n_f$ the proportion of free ions, $\tau_a$ and $\tau_d$ the ion pair association and dissociation times, respectively. One has $n_f + n_p = 1$ and the conductance is simply given by:
\begin{equation}
	G[n_f] = n_f G_\infty
\end{equation}
where $G_\infty$ is the conductance in the fully dissociated case. Overall, the system is able to remember the application of an electric field in its recent past over a timescale $\tau_a$, and Bjerrum pairs can be used to create memristors.

The above process cannot be achieved in bulk water, which fully dissociates all commonly used salts. However, as noted by Ref. \cite{robin_modeling_2021}, ionic interactions are greatly increased under confinement. This results in the formation of non-conducting Bjerrum pairs, sometimes to the point that there are no `free' ions left at thermal equilibrium. In this case, the conductance vanishes in absence of an electric field, $G(0) = 0$, and it can be shown that:
\begin{equation}
	G(E) \propto \ell_E^{-\alpha}
\end{equation}
with $\alpha > 1$ scaling like the strength of ionic interactions. In typical cases, one has $\alpha \sim 2$. In many experimental examples, however, a non-zero conductance remains even in the absence of voltage. Taking into account the fact that there may actually be a few free ions left at equilibrium, we then write:
\begin{equation}
	G(E) = G_0 + G_1 \left(\frac{|E|}{E_0}\right)^\alpha
\end{equation}
with $G_0 \ll G_1$.  This corresponds to equation $(1)$ of main text. The number of ion pairs play the role of an internal state variable governing the system's conductance. It should be noted, however, that the value of $G_0$ was found to vary from device to device (even with similar dimensions), although the general shape of the IV curves was preserved, see Fig. S11.

\subsection{Bipolar memristors -- Ionic rectification}
\label{ssec:Rect}
{
	The above mechanism, based on Wien effect, can only provide an explanation for the unipolar memristor effect, as only the absolute magnitude of voltage, and not its sign, governs the dynamics of ion pairs. Consequently, bipolar memristors must have a different origin, as they display different conductance states depending on the sign of applied voltage
	
	This observation is remindful of ionic rectification and nanofluidic diode (Fig.~S6). This phenomenon typically appears in systems that are spatially asymmetric, e.g. conical pores (geometrical asymmetry) \cite{siwy2002rectification}, channels connecting reservoirs of different salt concentrations (chemical asymmetry) \cite{picallo2013nanofluidic} or with an inhomogeneous surface charge (electrostatic asymmetry) \cite{karnik_rectification_2007}. In Ref. \cite{poggioli2019beyond}, the authors show that in all those cases the underlying mechanism is the same: the local value of the conductivity is increased where surface effects are the strongest (e.g., where the channel is the thinnest, salt concentration the lowest or surface charge the highest). Globally, the system then behaves like a PN junction: when current flow from the side of highest conductivity to that of lowest conductivity, ions accumulate inside the channel and conductance increases. Conversely, conductance is lowered if current flows from the side of lowest conductivity.
	
	For the sake of example, let us consider a 2D nanochannel with varying thickness $h(x)$ or surface charge $\Sigma(x)$, $x$ being the direction along the channel. The local value of salt concentration $c(x)$ will depend on $h(x)$, $\Sigma(x)$ and the salt concentration in the reservoirs $c_0$. The exact relation will depend on wether or not the system is in the Debye overlap regime. If the Debye length is large compared to the channel's thickness, then conductivity reads:
	\begin{equation}
		g(x) = 2 w \frac{e^2 D}{k_B T}\sqrt{(ec_0 h(x))^2 + \Sigma(x)^2}
	\end{equation}
	whereas in the opposite regime (Debye length small compared to $h$) it reads:
	\begin{equation}
		g(x) = 2 w \frac{e^2 D}{k_B T}(ec_0 h(x) + |\Sigma(x)|)
	\end{equation}
	with $w$ the channel's width, $D$ the diffusion coefficient of ions (assumed to be the same for both cations and anions), $e$ the elementary charge and $T$ temperature. 
	\rev{In particular, in the case of strong surface charges ($\Sigma \gg hec$), conductance is governed by the surface charge alone:
		\begin{equation}
			g \sim 2 w \frac{e^2 D}{k_B T}\Sigma
		\end{equation}
		In addition, carbon-based nanochannels like activated \rev{carbon} channels are known for their high hydrodynamic slip-length \cite{emmerich2022}. Due to the differential friction of ions and water molecules on the channel's walls, hydrodynamic slippage yields a correction to the conductance that scales like:
		\begin{equation}
			\Delta g \propto \frac{b \Sigma^2}{\eta}
		\end{equation}
		where $b$ is the slip length and $\eta$ the viscosity of water. In what follows, we will not discuss further the impact of slippage, but the two above equations show how even thin channels can reach a conductance of the order of $100 \,$nS thanks to their strong surface charge.}
	
	In all cases, conductivity can always be expressed in terms of the Dukhin number defined as:
	\begin{equation}
		\Du(x) = \frac{\Sigma(x)}{ec_0h(x)}
	\end{equation}
	and it increases with $\Du$ in all cases. Several regimes can now be identified:
	\begin{enumerate}
		\item $\Du \ll 1$ in the entire channel: bulk conduction dominates and the system conducts linearly; there is no ionic rectification.
		\item $\Du \gg 1$ in the entire channel: surface conduction dominates, and conductance is essentially given by counterions of the surface charge. As the number of counterions is fixed, there can be no accumulation nor depletion of ions in the channel, and no ionic rectification.
		\item $\Du \sim 1$ in at least parts of the channel, and is inhomogeneous: ionic rectification can occur.
	\end{enumerate}
	Quantitatively, Ref. \cite{poggioli2019beyond} computes the global conductance of a channel with a constant gradient of Dukhin number, in the absence of Debye overlap. The obtained ratio of the maximum conductance of the channel (for a strongly positive voltage) to its minimum conductance (for a strongly negative voltage) is given by:
	\begin{equation}
		\beta = \frac{G_\text{max}}{G_\text{min}} = \frac{1 + 2 \Du_{\min} } {1 + 2  \Du_{\max} }\frac{\Du_{\max} }{\Du_{\min} }
	\end{equation}
	where $\Du_{\min}$ and $\Du_{\max}$ are the minimum and maximum values of the Dukhin number, respectively.
	
	The bipolar memristor effect was typically observed in activated \rev{carbon} channels at low salt concentration. Typical parameters are $\Sigma \sim 0.1\,\si{C/m^2}$, $c_0 \sim 1\,$mM and $h \sim 10\,$nm, corresponding to $\Du \sim 10^2$. Yet, we observed ionic rectification with $\beta \sim 1-5$, corresponding to $\Du_{\min} \sim 1$ in the above equation (assuming $\Du_{\max} \gg 1$). Such variations of the Dukhin number over several orders of magnitude across the channel cannot be accounted for by fluctuations of the channel's height or surface charge, especially in the case of pristine MoS$_2$ channels, which are atomically smooth throughout. Therefore, it seems more reasonable to attribute the ionic rectification observed in 2D nanochannels to entry effects: the nanofluidic device is deposited on a SiN$_x$ membrane, which is typically much thicker than the device itself (500\,nm versus $\sim$50\,nm, respectively). The hole in the bottom layer thus could act as an additional channel in series with the nanofluidic device, with similar surface charge but much larger spatial extension ($\sim 1\,\si{\micro m}$), yielding $\Du \sim 1$.
	
	This asymmetry in entry effects could be the source of an accumulation or a depletion of ions inside the channel, depending on the sign of voltage.
}

\subsection{Minimal model of a nanochannel with long-term memory}
In this section, we complement the above analysis by showing how long-term memory can emerge from interfacial processes in 2D nanochannels. We first show how the rectification mechanism can be slowed down considerably if there is adsorption of ions on the channel's walls. This allows us to derive a minimal model with analytical solutions, which can then be extended to more complex cases, like that of the second Wien effect. However, we start by taking a step back to analyze the nature of nanofluidic memory.

\subsubsection{What does it mean for a nanochannel to have memory?}
\label{ssec:DiffMem}
In this section, we discuss what we really measure in a memristor experiment, and how we can quantify a system's memory, with the help of a simplistic advection-diffusion model.

We first focus on channels exhibiting ionic rectification. Rather than taking the model presented in previous section, with geometrical asymmetry, we simplify the problem even further. We assume that the channel has a concentration $c$ of ions, and is connected to two reservoirs. We introduce an ad hoc asymmetry by assuming that the reservoir on the left of the channel has concentration $c_L$ and the reservoir on the right $c_R$, to mimick ionic rectification while keeping mathematical complexity at a minimal level. Lastly, we assume that there is an external forcing $f$ driving particles from the left to the right. For the sake of simplicity, we work with units such that the diffusion time across the channel, $L^2/D$, is equal to $1/2$.  

In a continuous problem, the particle flux would be:
\begin{equation}
	j(x,t) = -\frac 1 2\partial_x c + f(t) c(x,t) = j_\text{diff} + j_\text{adv}
\end{equation}
Note that the diffusive part $j_\text{diff}$ is an artefact{: it exists because we replaced a spatial asymmetry by a concentration gradient, resulting in a global diffusive flux}; it is irrelevant in actual experiments. The real physical quantity of interest is therefore $j_\text{adv}$. In practice, we measure this flux at the electrodes, i.e. at both ends of the channel, without access to the full spatial dependence of $j$. {In a time-independent problem with $f(t) = f_0$ at all times, flux is conserved and this does not matter; we obtain $j_\text{measured} = f_0 \times \mean {c}_{\infty,f_0}$, where $\mean {c}_{\infty,f_0}$ is the spatial average of $c$ when the channel is subject for a forcing $f_0$ for a very long time. In the quasistatic limit, one would replace $f_0$ by a slow-varying forcing $f(t)$:
	\begin{equation}
		j_\text{quasistatic} = f(t) \times \mean{c}_{\infty,f(t)}
	\end{equation}
	where $\mean{c}_{\infty,f(t)}$} is now the spatial average of $c$ when the channel is subject for a forcing ``frozen'' at a specific value $f(t)$. A memory effect is any deviation from the above equation; it translates the fact that the forcing is varying faster than the time needed to equilibrate the system quasistatically. In other words, memory is stored in the number of particles that can contribute to conduction.

However, when the system is not in the quasistatic limit, flux is not conserved across the system. It is then hard to make exact sense of what is being actually measured at the electrodes; in the following, we admit we may still assume we measure a spatial average of the advection flux, but that this average is now \textit{instantaneous}:
\begin{equation}
	j_\text{measured} \simeq f(t) \mean {c(x,t)}_x = f(t) \frac 1 L \int c(x,t) \, \di x
\end{equation}
Assuming the forcing is sinusoidal $f(t) = f_0 \cos \omega t$, one can then compute the area of the hysteretic loop in the conductance-voltage curve:
\begin{equation}
	\mathcal A(\omega) = \left| \oint \mean {c(x,t)}_x \, \di f\right| = \int_0^{2\pi/\omega} \omega f_0 \mean {c(x,t)}_x \sin \omega t \, \di t
\end{equation}
The bigger $\mathcal A$ is, the more memory the system has of its recent past. We thus define the memory timescale as $\tau_m = 2 \pi/\omega_m$ such that $\mathcal{A}(\omega_m)$ is maximum.

There is, however, no simple way to solve the advection-diffusion equation under periodic forcing, even in 1D, so we simplify the problem further by writing an approximate equation for the mean $\mean c$ only, see Fig. S7A:
\begin{equation}
	\dot {\mean c} + \mean c \simeq \frac{c_L + c_R}{2} + f(t) \frac{c_L - c_R}{2}
\end{equation}
This model is summed up in Fig. S7. In all what follows, we drop the $\mean \cdot$ for the sake of simplicity. This yields:
\begin{equation}
	c(t) = \frac{c_L + c_R}{2} + f_0 \frac{c_L - c_R}{2}\frac{\cos \omega t + \omega \sin \omega t}{1 + \omega^2} \label{eqn:SolRect}
\end{equation}
so that the loop area is:
\begin{equation}
	\mathcal A (\omega) =f_0 \frac{c_L - c_R}{2} \frac{\omega}{\omega^2 + 1}
\end{equation}
This yields $\tau_m = 2 \pi$, or, in dimensional terms:
\begin{equation}
	\tau_m = \pi \frac{L^2}{D}
\end{equation}
This is perfectly intuitive: since information is encoded in the particle number, it cannot be retained for more than the diffusion time. However, nanochannels have $L < 10 \, \si{\micro \meter}$, meaning that the diffusion time cannot exceed a second, contrary to what is observed in experiments ($\tau_m \sim 1$ hour). 

Before we move on to a slightly modified version of this model to account for this, let us make the following remark. The quasistatic solution to the above problem is:
\begin{equation}
	c_\text{qs}[f] = \frac{c_L + c_R}{2} + f\frac{c_L - c_R}{2}
\end{equation}
The instantaneous solution we obtained can be rewritten into the following form:
\begin{equation}
	c(t) = \int_0^{+\infty} c_\text{qs}[f(t-s)] \frac{e^{-s/\tau}}{\tau} \, \di s \label{eqn:ansatz}
\end{equation}
where $\tau = 1$, equal to $\tau_m$ up to a factor of order unity. This equation will allow us to model more complex situation, where analytical solutions do not exist.

\subsubsection{Adsorption-desorption model}
\label{ssec:SurfMem}
To complement the above model, which does not account for long memory times observed in experiments, we consider the possibility of ion adsorbing on the channel's walls, and denote the number of adsorbed particles by $\sigma$. Introducing $k$ and $\lambda$, the adsorption and desorption rates, respectively, we obtain (see Fig. S7B):
\begin{align}
	\dot c + c &= \frac{c_L + c_R} 2 + f(t) \frac{c_L - c_R}{2} - kc + \lambda \sigma\\
	\dot \sigma &= kc - \lambda \sigma
\end{align}
It again can be solved analytically:
\begin{align}
	c(t) &= \frac{c_L + c_R} 2 + f_0 \frac{c_L-c_R}{2}\frac{\left[\lambda^2 + \left(1 + k\right)\omega^2\right] \cos \omega t + \omega \left[\lambda \left(k + \lambda\right) + \omega^2\right] \sin \omega t}{\lambda ^2 + \left[\left(1 + k\right)^2 + 2 k \lambda + \lambda^2\right] \omega^2 + \omega^4} \label{eqn:SolRectSurf}\\
	\mathcal A (\omega) &= f_0 \frac{c_L-c_R}{2}\frac{\omega \left[\lambda \left(k + \lambda\right) + \omega^2\right]}{\lambda ^2 + \left[\left(1 + k\right)^2 + 2 k \lambda + \lambda^2\right] \omega^2 + \omega^4}
\end{align}
The memory timescale is then given by:
\begin{align}
	&\left\{\lambda \left(k + \lambda \right) + 3\omega^2 \right\} \left[ \lambda^2 + \left\{ \left(1 + k\right)^2 + 2 k \lambda + \lambda^2\right\} \omega^2 + \omega^4\right]\\
	&...= \omega\left\{\lambda \left(k + \lambda\right) + \omega^2\right\} \left[2\left\{ \left(1 + k\right)^2 + k \lambda + \lambda^2\right\} \omega + 4\omega^3\right]\
\end{align}
There is no closed-form solution to this last equation, but we can extract an approximate solution when surface effects strongly dominate ($k \gg \lambda$ and $k \gg 1$):
\begin{equation}
	\omega_m \sim \frac{\lambda}{k} \ll 1
\end{equation}
In other words, the memory time reads in this case:
\begin{equation}
	\tau_m = 2 \pi \frac{\sigma_\infty}{c_\infty} \frac{L^2}{D}
\end{equation}
where $\sigma_\infty$ and $c_\infty$ are the values of $\sigma$ and $c$ at chemical equilibrium, respectively. The ratio $\sigma_\infty/{c_\infty}$ appearing in the above equation is the analogue of the Dukhin number \cite{kavokine_fluids_2021}, which measures the importance of the surface charge of a channel with respect to the bulk concentration in ions. This number can reach several hundreds, so $\tau_m$ can be of the order of several minutes.

Qualitatively, the above equation can be recovered from a semi-quantitative argument as follows. The memory time is given by the maximum time a particle can stay within the channel. It will reach one of the reservoirs if left free for more than $L^2/D$, by randomly diffusing along the channel. However, every $1/k \ll 1$, the particle is adsorbed and stops moving, only liberated after a time $1/\lambda$. Along its course through the channel, there will therefore be $kL^2/D \gg 1$ such events of duration $1/\lambda$. The total time spent inside the channel is the sum of the ``travelling time'' and ``resting time'':
\begin{equation}
	\tau_m \sim \frac{1}{\lambda}\frac{kL^2}{D} + \frac{L^2}{D} \sim \frac{k}{\lambda}\frac{L^2}{D}
\end{equation}
The last approximation holds since we assumed that $k\gg\lambda$, such that the total resting time is much larger than the travelling time.

To complement the above minimal model, we also computed a numerical solution of the 1D advection-diffusion under periodic forcing and with adsorption, yielding similar results (not reported here).

\subsubsection{Wien effect}

In the previous model, long-term memory emerges from a stop-and-go mechanism of ions being adsorbed and desorbed. However, this process is more general than the specific physics of adsorption, and we can write a similar system taking into account a Wien effect mechanism:
\begin{align}
	\dot c + c &= c_0 - \frac{1}{\tau_a}c^2 + \frac{1}{\tau_d[f(t)]} p\\
	\dot p &= \frac{1}{\tau_a}c^2 - \frac{1}{\tau_d[f(t)]} p
\end{align}
Here, $c$ is the concentration of ions that can contribute to conduction (free ions or polyelectrolytes), while $p$ represents the concentration of pairs. This system is, however, non-linear in both $c$ and $f$, and as such admits no analytical solution, but is qualitatively similar to the previous linear case. Rather than looking for an approximate solution of this already simplistic model, we instead use the ansatz derived above.

\subsubsection{A simple ansatz - Determination of the memory time}

Here, we detail how we fitted the curves presented in Figure 3 of main text. As per sections \ref{ssec:Wien} and \ref{ssec:Rect}, we have two models describing the conductance of a 2D nanochannel under a \textit{constant} electrical forcing. As shown by equation \eqref{eqn:ansatz}, this conductance becomes in the timevarying-case:
\begin{equation}
	G(t) = \int_0^{+\infty} G_\text{qs}[\Delta V(t-s)] \frac{e^{-s/\tau}}{\tau} \, \di s 
\end{equation}
where $G_\text{qs}$ is the conductance in the stationary case (i.e. if the voltage was `frozen' at the value $\Delta V(t-s)$ for an infinite amount of time). Depending on cases, we use the following expressions:
\begin{align}
	G_\text{qs}(\Delta V) &= G_1 \Delta V^\alpha, \quad  \alpha = 2-3 \quad \text{(Wien effect)}\\
	G_\text{qs}(\Delta V) &= \begin{cases}
		G_+, & \text{ if } \Delta V >0,\\ 
		G_-, &  \text{ if } \Delta V <0
	\end{cases} \quad \text{(Ionic rectification)}
\end{align}
We first use these models to extract the quasistatic limit of experimental curves, and we then ``turn on'' memory effects by plugging the chosen model into equation \eqref{eqn:ansatz}. To assess the robustness of our approach, we can determine the memory time through two methods:
\begin{itemize}
	\item By using $\tau$ as a fitting parameter in equation \eqref{eqn:ansatz}.
	\item By noticing that, upon correct renormalization (see below), the area of the loop in the IV curve should take the form:
	\begin{equation}
		\mathcal A^*(\omega) \simeq K\frac{\omega \tau_m}{1 + \omega^2 \tau_m^2}
	\end{equation}
	with a theoretical value $K = 1$ in the adsorption-desorption model, and using $\tau_m$ as a fitting parameter in this last equation. In the case of the pairing-unpairing model, we cannot derive the exact expression of $\mathcal A^*$, but we use the above expression as an approximative ansatz, with $K$ as a fitting parameter.
\end{itemize}
Both quantities $\tau$ and $\tau_m$ can be interpreted as memory timescales, being equal up to a factor $\pi$ in the minimal model. In practice, both methods yield similar results ($\tau \sim \tau_m \sim 100 \, \si{s}$, see Figure 4 from main text), but we believe the second one ($\tau_m$) to be more robust as it is determined using more experimental data points obtained using several frequencies.

Let us now detail how we normalize the loop area. Memristive devices cycle between different conductance states. The largest loop that could possibly be observed would be in the case where the device switches abruptly between the lowest and the highest conductance states, $G_\text{off}$ and $G_\text{on}$, whenever it reaches $\Delta V = \pm V_0$ (for bipolar memristors) or $\Delta V = 0$ and $|\Delta V| = V_0$ (for unipolar memristors). The conductance-voltage curve then takes the shape of a rectangle of size $2 V_0 \times (G_\text{on} - G_\text{off})$. The IV curve then takes the shape of two triangles of total area $V_0^2 \times (G_\text{on} - G_\text{off})$, which we therefore use for normalization. For each case, we determine $G_\text{off}$ and $G_\text{on}$ graphically using data with the lowest frequency for a given device and salt concentration. In the case of devices with mixed unipolar-bipolar behavior, we determined both extremal conductance for positive and negative voltage, and then used $0.5 V_0^2 \times (G_{\text{on},+} - G_{\text{off},+}) + 0.5 V_0^2 \times (G_{\text{on},-} - G_{\text{off},-})$ as normalization. This normalization process is summed up in Fig. S8.

{
	\subsection{Mixed mechanisms and shape of the IV curve}
	
	The two mechanisms detailed above are not mutually exclusive, as some nanofluidic systems can display both ionic rectification and allow the formation of ionic pairs. In particular, pristine MoS$_2$ were found to display both behaviour depending on experimental conditions (Fig. S10). Additionally, some of our experimental data do not fall in either of the two phenomenologies: they displayed two crossing points in their IV curve, for example.
	
	To shed light on these `mixed' cases, we designed a simple model of a memristor that is intermediate between unipolar and bipolar cases:
	\begin{align}
		I(t) &= G(t) \times V(t)\\
		G(t) &= \int_{0}^{+\infty} G_\text{qs}(V(t-s)) \frac{e^{-s/\tau}}{\tau}\\
		G_\text{qs}(v) &= G_\text{unipolar}(1 + \alpha v^2) + G_\text{bipolar}(1 + \beta v)
	\end{align}
	with $I$ the ionic current, $V$ voltage, $G$ the instantaneous conductance, $G_\text{qs}$ the quasistatic conductance and $\tau$, $G_\text{unipolar}$, $G_\text{bipolar}$, $\alpha$ and $\beta$ some constants. For different values of the parameters, this model can be purely unipolar, purely bipolar, or intermediate between the two. Various examples of IV and GV curves for different values of parameters are shown on Fig. S17, along an example of experimental data displaying this mixed behaviour. We note that there exists a ``critical point" such that the GV curve has no crossing point; however, it develops a cusp, which transforms into a crossing point when parameters are varied.
}

\section{Additional experimental data}

\subsection{Pristine MoS$_2$ channels}

We provide additional data to characterize the memristor effect in pristine MoS$_2$ channels, which is found to be robust when we vary salt concentration, channel height and the chemical nature of the electrolyte.

On Fig. S9, we show that the memristor effect can be observed in pristine MoS$_2$ channels regardless of the electrolyte used (KCl, NaCl, LiCl, CaCl$_2$, NiSO$_4$), asserting its robustness. Despite variations in conductance, all these curves display the same non-linear general shape remindful of the Wien effect.

On Fig. S10, we provide additional data for the memristor effect in bilayer pristine MoS$_2$ channels with potassium chloride solutions of various concentrations. In particular, we notice that the system behaves like a bipolar memristor at low salt concentration.

On Fig. S11, we show the evolution of loops in IV curve at fixed frequency and salt concentration, but with increasing channel height. We find that the effect is most visible in thin channels, and that the loop collapses to a straight line in larger channels. This shows that memory effects can only be observed if confinement is sufficiently strong so that ion-ion interactions are enhanced, allowing the formation of ion pairs.

Lastly, Fig. S12 shows the determination of the memory time of three different devices (and two {salt} concentrations) from the evolution of the loop area with frequency. We obtain values between $50$ and $400 \, \si{s}$.

\subsection{Activated \rev{carbon} channels}

Additional data characterizing the influence of salt concentration on memristive effects in activated \rev{carbon} channels are presented in Fig. S13. They notably include raw data for Fig. 2A of main text. We observe that {salt} concentration variations have little effect on the memory of thinner channels (Fig. S13A-B), while this influence is more visible for devices above $10 \, \si{nm}$ in thickness. This notably shows that, in more confined systems, interfacial processes are stronger than bulk effects.

In addition, the memristor effect can also be observed with salts other than CaCl$_2$, as shown in Fig. S14 for KCl and AlCl$_3$.

We also provide additional data regarding the effect of voltage frequency (Fig. S15), corresponding to Fig. 4C of main text. These data allow us to compute the memory time of each activated system as the inverse of the frequency such that the loop in the IV curve is the largest. We obtain values spanning from $50$ to $400 \, \si{s}$, in a similar range as pristine MoS$_2$ channels. Such variations may be explained by the variability of the surface state of activated carbon channels following the etching in low pressure water vapor.

\section{Detailed list of supplementary figures}

We summarize below the content of all supplementary figures:

\begin{itemize}
	\item Fig. S1: Nanofabrication of pristine MoS$_2$ channels.
	\item Fig. S2: Nanofabrication of activated \rev{carbon} channels.
	\item Fig. S3: Details of the long-term potentation algorithm.
	\item Fig. S4: Details of the Hebbian learning algorithm
	\item Fig. S5: Additional data for long-term potentiation and Hebbian learning.
	\item Fig. S6: Sources of ionic rectification in nanofluidics.
	\item Fig. S7: Minimal model of nanofluidic memory.
	\item Fig. S8: Normalization process of loop area in memristive GV curves.
	\item Fig. S9: Additional data: pristine MoS$_2$ channels with different salt types
	\item Fig. S10: Additional data: pristine MoS$_2$ channels with different salt concentrations.
	\item Fig. S11: Additional data: pristine MoS$_2$ channels of different heights.
	\item Fig. S12: Determination of the memory time of pristine MoS$_2$ channels.
	\item Fig. S13: Additional data: activated carbon channels with different salt concentrations and different heights.
	\item Fig. S14: Activated carbon channels with different salt types.
	\item Fig. S15: Determination of the memory time of activated carbon channels.
	\item Fig. S16: Influence of pH.
	\item Fig. S17: `Mixed' memristor types (theoretical model and experimental example).
\end{itemize}

\clearpage

\begin{figure}[h!]
	\centering
	\includegraphics[width=\textwidth]{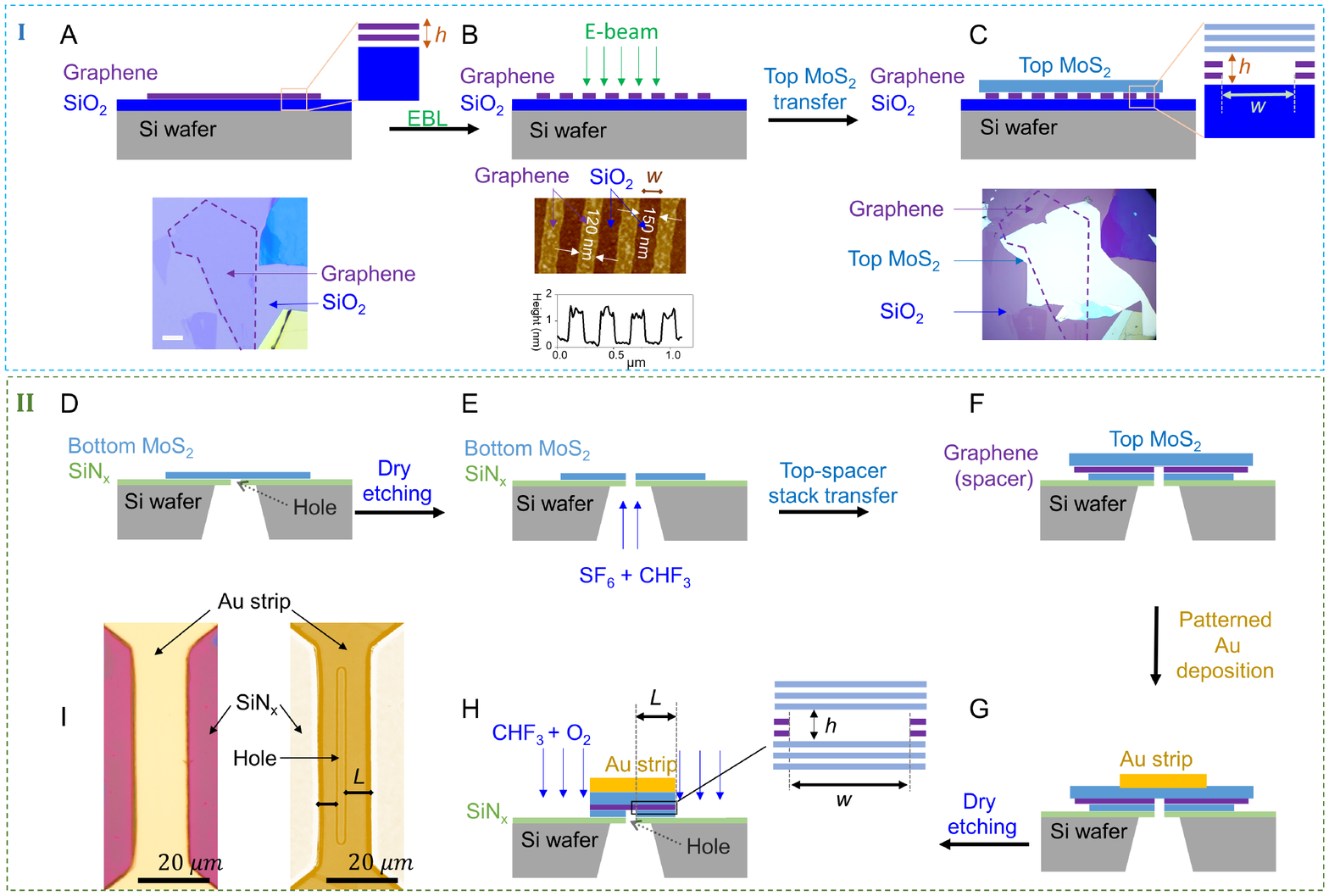}
	\caption*{\label{SI}Fig. S1. \textbf{Schematic flow-chart in cross-sectional view and corresponding optical and AFM images.} \textbf{Step I:} Preparation of the top-spacer layers on silicon/silicon oxide (SiO$_2$) wafer. \textbf{A} Graphene is mechanically exfoliated; the thickness of this graphene flake will determine the height ($h$) of the channel. Bottom panel: Optical image of a 3-layer graphene spacer. \textbf{B} Patterning of the spacer using electron beam lithography (EBL) and etching into parallel strips. Bottom panel: atomic force microscopy (AFM) image and height profile of the patterned graphene-spacer ($h \sim$ 1.2 nm and $w \sim$ 150 nm). \textbf{C} Transfer of MoS2 flake as a top layer over graphene spacer. Bottom panel: Optical image of the top-spacer stack. \textbf{Step II:} Assembly of the tri-crystal (top-spacer-bottom) stack on silicon/silicon nitride (SiN$_x$) wafer. \textbf{D} Transfer of a MoS$_2$ flake onto SiN$_x$ membrane bearing a hole ($\sim 3 \times 50 \, \si{\micro m}$), to serve as bottom wall of the channel. \textbf{E} Dry etching of the MoS$_2$ bottom layer from the back of the SiNx. \textbf{F} Transfer of the top (MoS$_2$)-spacer (graphene) stack prepared in C over the bottom MoS$2$ prepared in E. \textbf{G} Patterned gold (Au) deposition over the tri-crystal stack. \textbf{H} With the Au strip as a mask to protect the underneath channels, the surrounding regions are etched away. The gold strip thus determines the channel length ($L$). \textbf{I} Optical images (left: reflection mode, right: transmission mode) of the final channel device. The tricrystal device is underneath the gold strip on the SiN$_x$ membrane. Scale bar in all images represents $20 \, \si{\micro \meter}$.}
\end{figure}

\begin{figure}[h!]
	\centering
	\includegraphics[width=\textwidth]{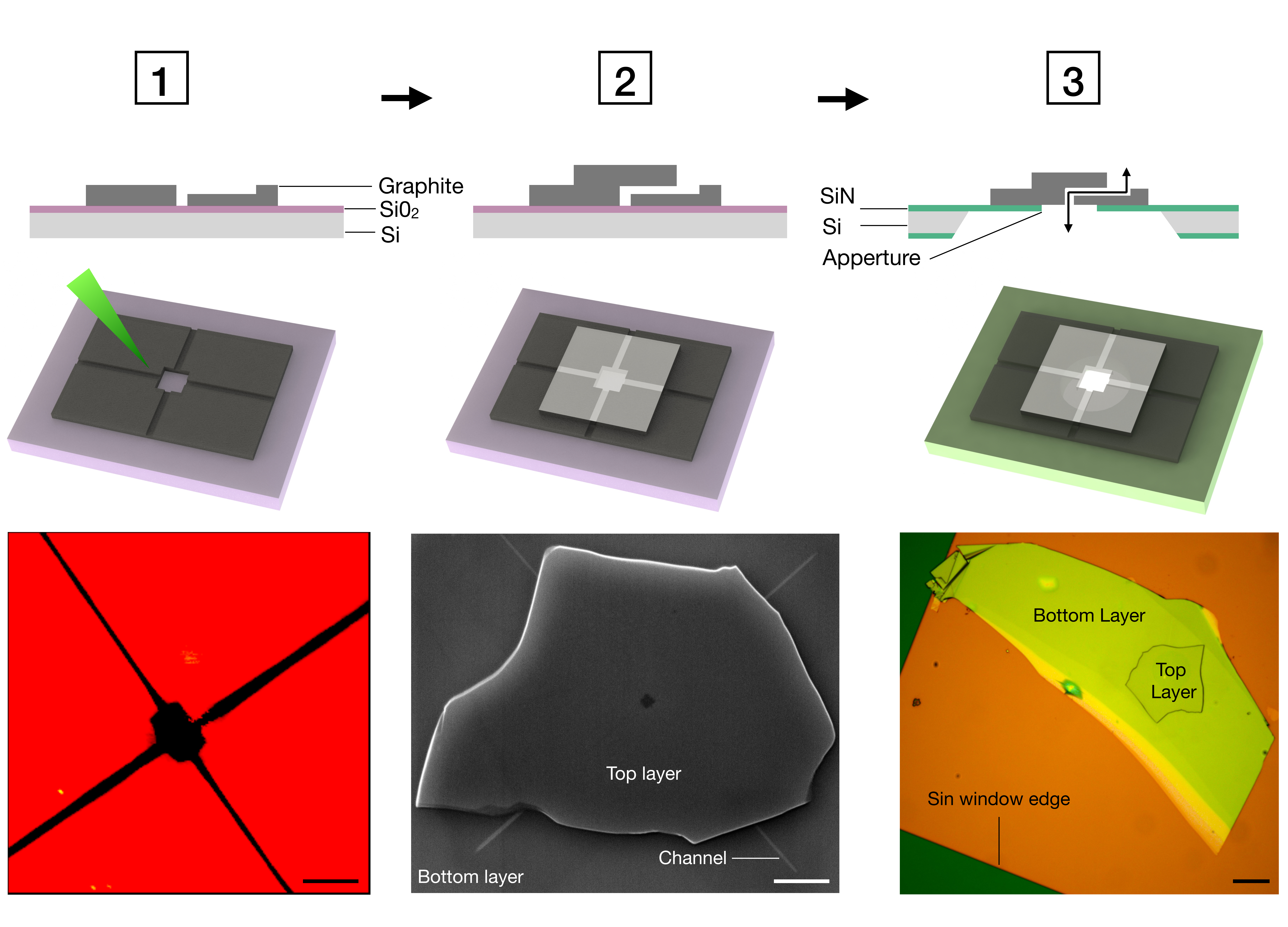}
	\caption*{\label{Nanofab_activated}Fig. S2. \textbf{Fabrication of activated carbon channels. Step 1: Patterning of the bottom layer.} Up:  2D side view of a graphite bottom layer crystal after the patterning on a Si/SiO$_2$ substrate. Middle: 3D view of the patterning process. A square-shaped hole and four trenches connected to the hole are represented. Bottom layer graphite is dark grey and the Si/SiO$_2$ substrate is light pink. The electron flux, represented as a green tip, enables selective removal of matter. Down: AFM image of the bottom layer after etching. Scale bar represents $1 \, \si{\micro m}$. \textbf{Step 2: Dry transfer of the top layer.} Up: 2D side view. A top layer crystal is added above the bottom layer. Middle: 3D view of the device after the transfer of the top layer, represented in glassy transparent grey. Down: SEM image of a device at that stage. Four channels are visible in white. The bottom layer hole remains visible through the top layer. Scale bar represents $5 \, \si{\micro m}$. \textbf{Step 3: Wet transfer on the Si/SiN$_x$ membrane.} Up: 2D side view. Middle: 3D view, with the SiN membrane in green. The circular aperture in the SiN$_x$ membrane is visible by transparency. Down: Optical microscope image of a finished device. Scale bar represents $10 \, \si{\micro m}$. }
\end{figure}

\begin{figure}[h!]
	\centering
	\includegraphics[width=\textwidth]{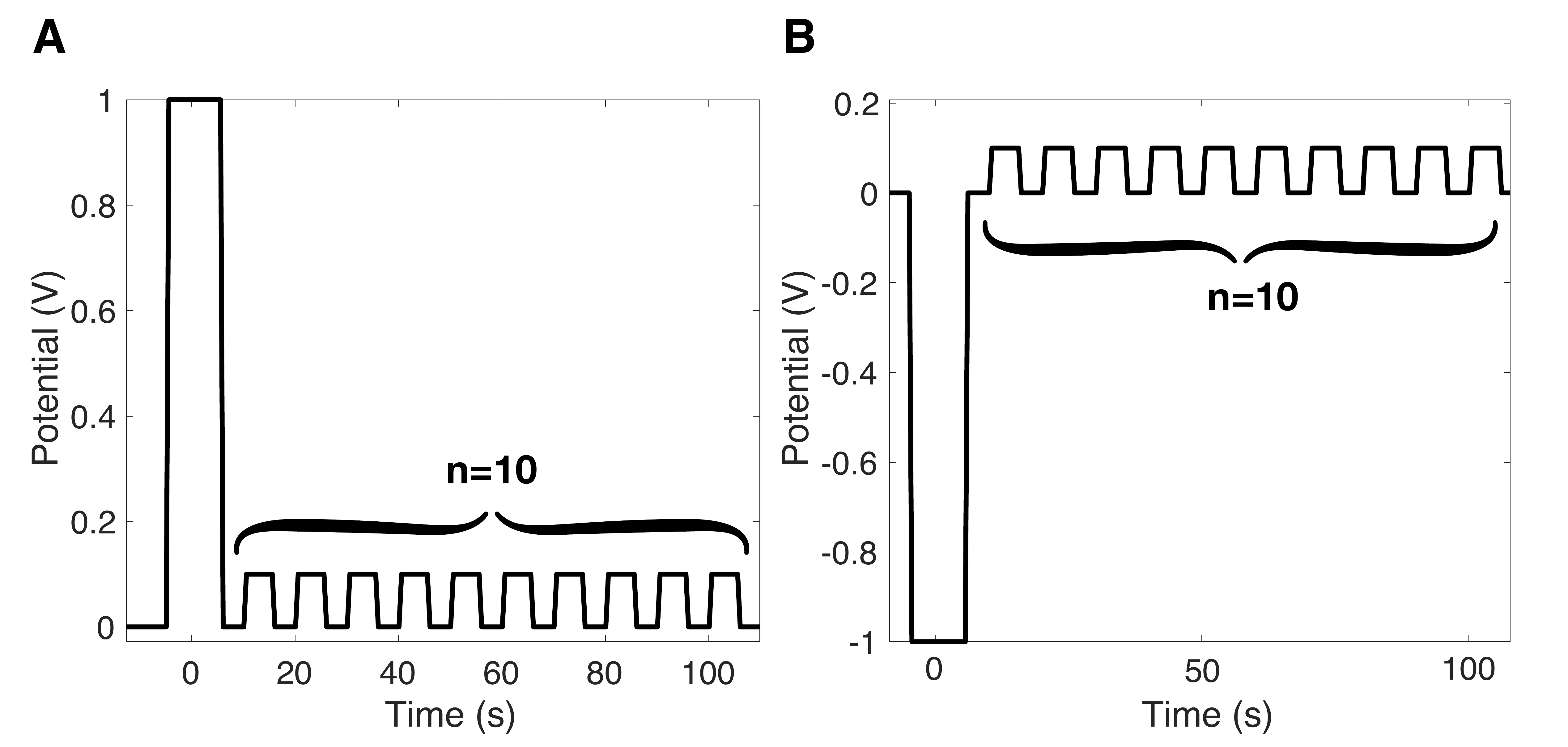}
	\caption*{\label{SI_Synapse}Fig. S3. \textbf{Long-term modification of a nanochannel conductance using voltage pulses}. (\textbf{A}) A `write' pulse ($+1 \, \si{V}$, $10 \, \si{s}$), followed by ten `read' pulses ($+0.1 \, \si{V}$, $10 \, \si{s}$) to study the relaxation of the conductance. (\textbf{B}) An `erase' pulse ($-1 \, \si{V}$, $10 \, \si{s}$), followed by ten `read' pulses ($+0.1 \, \si{V}$, $10 \, \si{s}$).}
\end{figure}

\begin{figure}[h!]
	\centering
	\includegraphics[width=\textwidth]{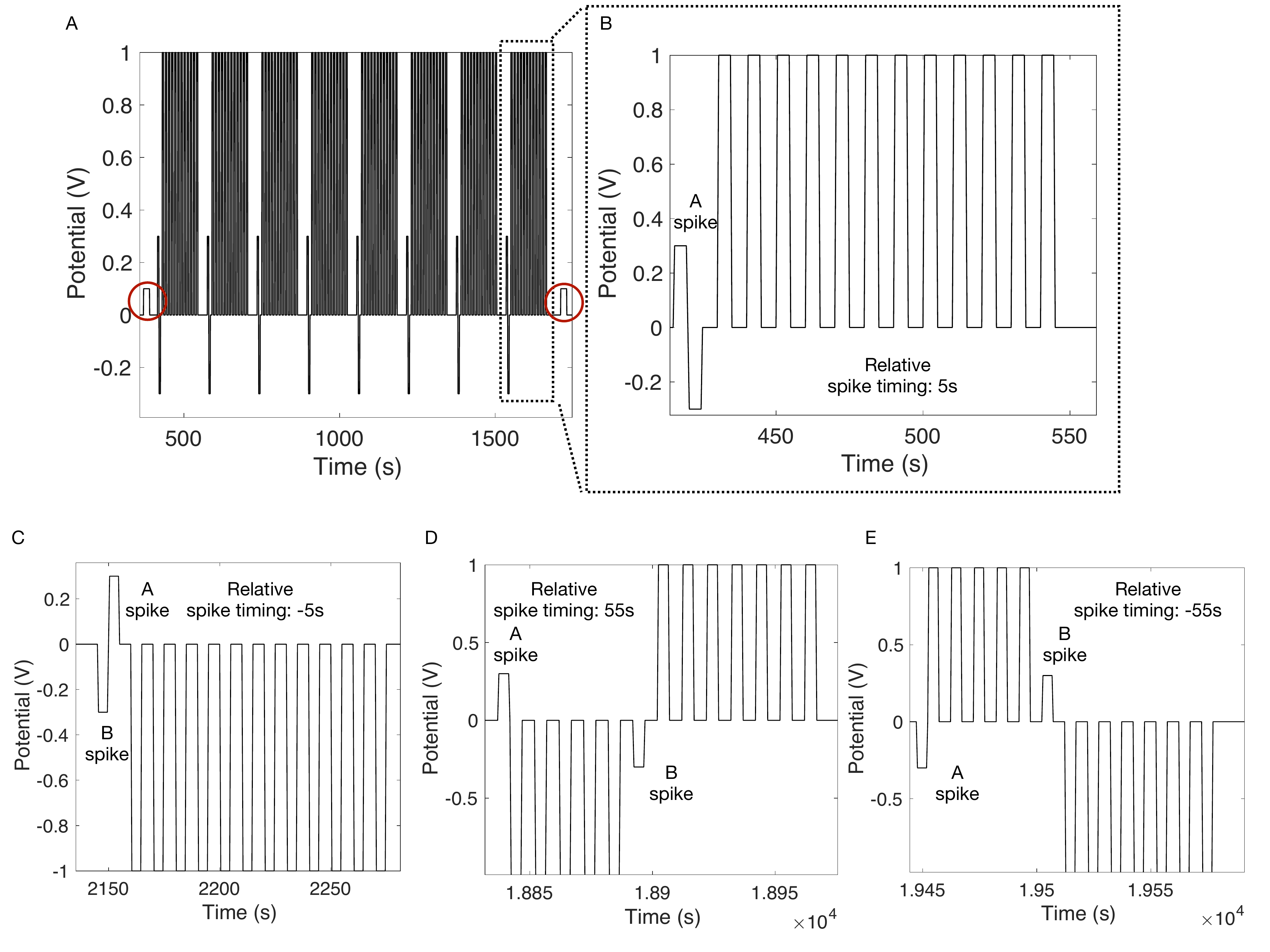}
	\caption*{\label{Hebb's rule}Fig. S4. \textbf{Algorithm for the implementation of Hebb's rule with activated carbon channels}. (\textbf{A}) Voltage input emulating 8 successive activation of a pre-synaptic and a post-synaptic neuron with a relative spike timing $\Delta t = 10 \, \si{s}$. The conductance is read before and after with low amplitude `read' pulses, highlighted by red circles. (\textbf{B} to \textbf{E}) Examples of input voltage for various relative spike timings of the two neurons.}
\end{figure}

\begin{figure}[h!]
	\centering
	\includegraphics[width=\textwidth]{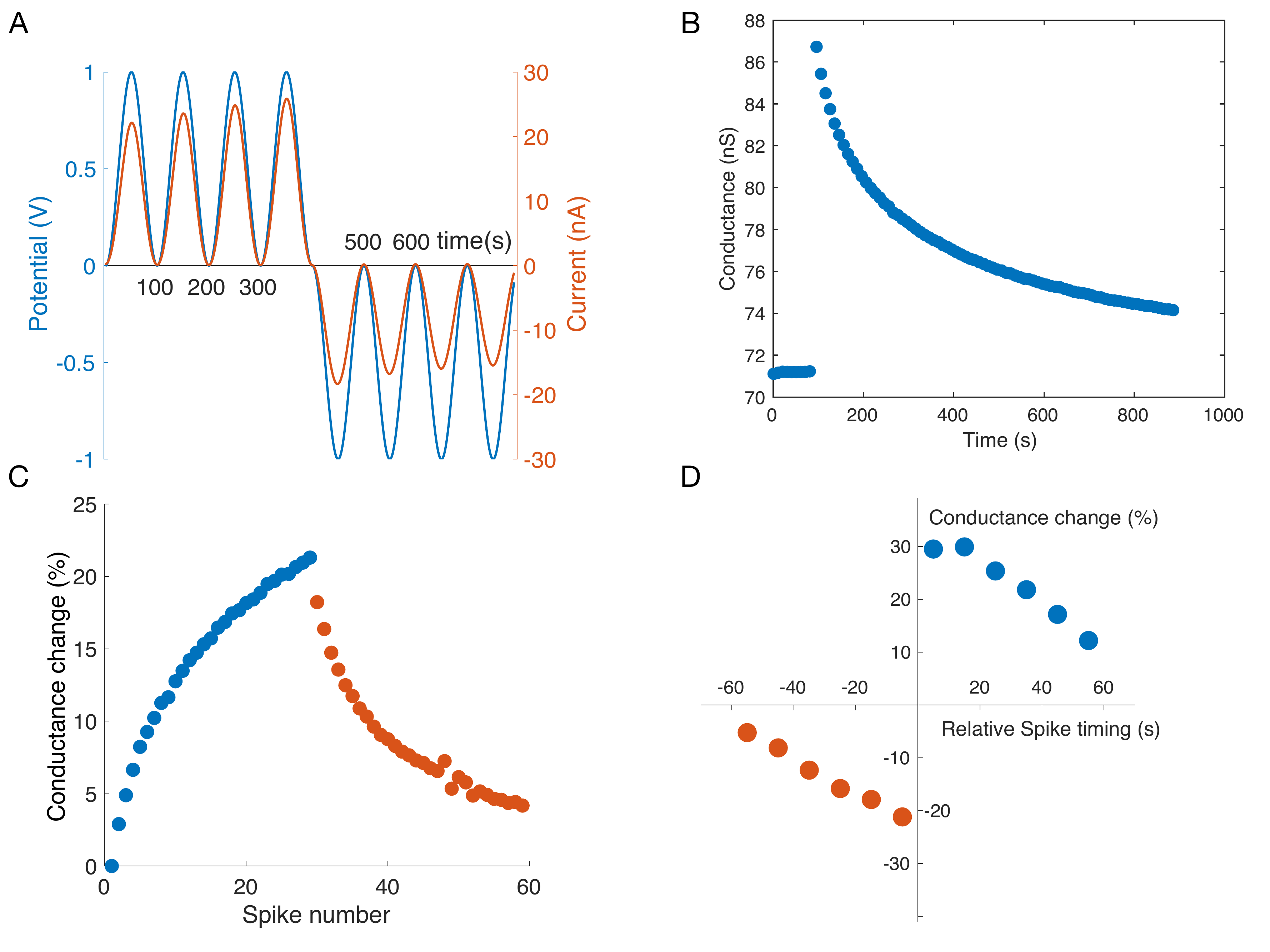}
	\caption*{\label{write_read_relax}Fig. S5. \textbf{Additional data for the implementation of neuromorphic computing with activated carbon channels.} The salt used is CaCl$_2$, $1 \, \si{mM}$. (\textbf{A}) Evolution of the ionic current (red) under voltage pulses of constant {polarity} (blue) (see Fig. 5A of main text). (\textbf{B}) Conductance change following a positive voltage pulse of $1 \, \si V$ in amplitude and a duration of $20\, \si{s}$ (see Fig. 5B of main text). The conductance is read with an alternating square voltage of $0.1 \, \si{V}$ in amplitude and a period of $20 \, \si{s}$. (\textbf{C}) Revsersible, long-term modification of a nanochannel’s conductance. 30 write spikes (+1 V, 10 s) are applied, followed by 30 erase spikes (-1 V, 10 s) which bring back the system to its initial state. Between each spike, the conductance is let to stabilize during two minutes and is then measured with a read pulse (0.1 V, 5 s), see Fig. 5C of main text. (\textbf{D}) Conductance change after 8 successive activations of the two neurons, in percentage of the initial conductance and as function of the relative activation timing of the simulated neurons (see Fig. 6D of main text).}
\end{figure}

\begin{figure}
	\centering
	\includegraphics[width=1\linewidth]{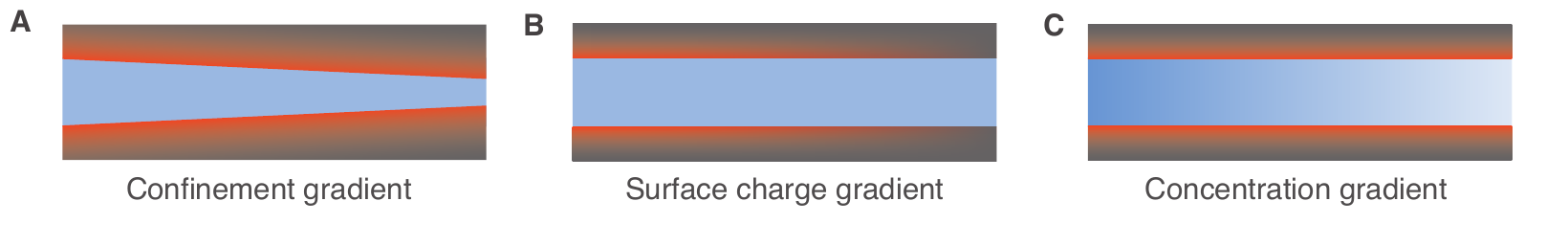}
	\caption*{{Fig. S6. \textbf{Possible sources of ionic rectification.} Nanochannel with variable (\textbf A) height, (\textbf B) surface charge or (\textbf C) concentration gradient. In all three cases, a gradient of Dukhin number is established across the channel, resulting in a conductivity gradient and ionic rectification.}}
\end{figure}

\begin{figure}
	\centering
	\includegraphics[width=1\linewidth]{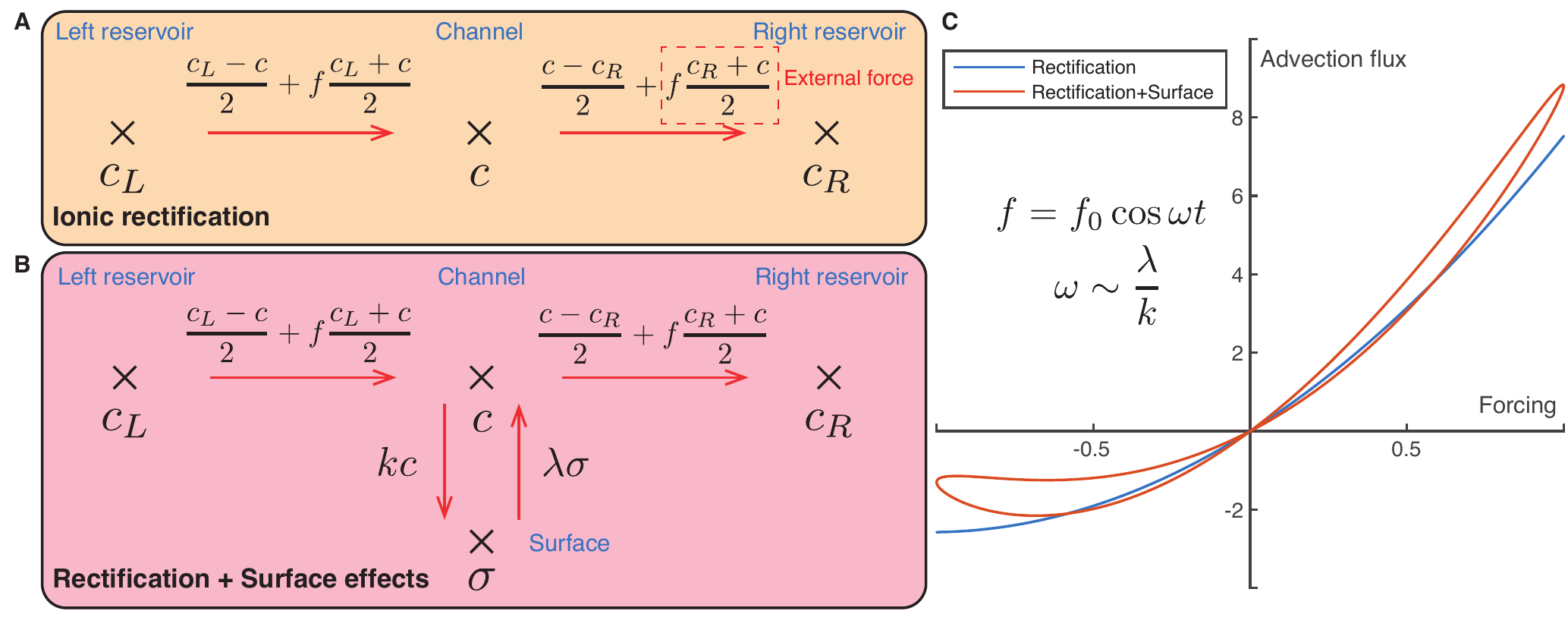}
	\caption*{Fig. S7. \textbf{Minimal model of nanofluidic memory.} (\textbf A) Minimal model of purely diffusive memory. The nanochannel is modeled by a single point that exchanges particles with two reservoirs. The concentration difference between reservoirs plays the role of geometrical asymmetry. Memory is stored in the concentration inside the channel $c$, and the memory time is the diffusion timescale. (\textbf{B}) Same minimal model, but with particle adsorption on the channel's walls. Transport is now limited by a stop-and-go mechanism of particles adsorbing and desorbing from the walls, giving rise to a memory time orders of magnitude larger than diffusion. (\textbf{C}) Memristor effect in the minimal model, as shown by the loop in the IV curve, in dimensionless units. The blue curve corresponds to the model described in panel A (section \ref{ssec:DiffMem}), and the red one in panel B (section \ref{ssec:SurfMem}). Memory effects are visible at low frequency only when surface effects are taken into account.}
\end{figure}

\begin{figure}
	\centering
	\includegraphics[width=1\linewidth]{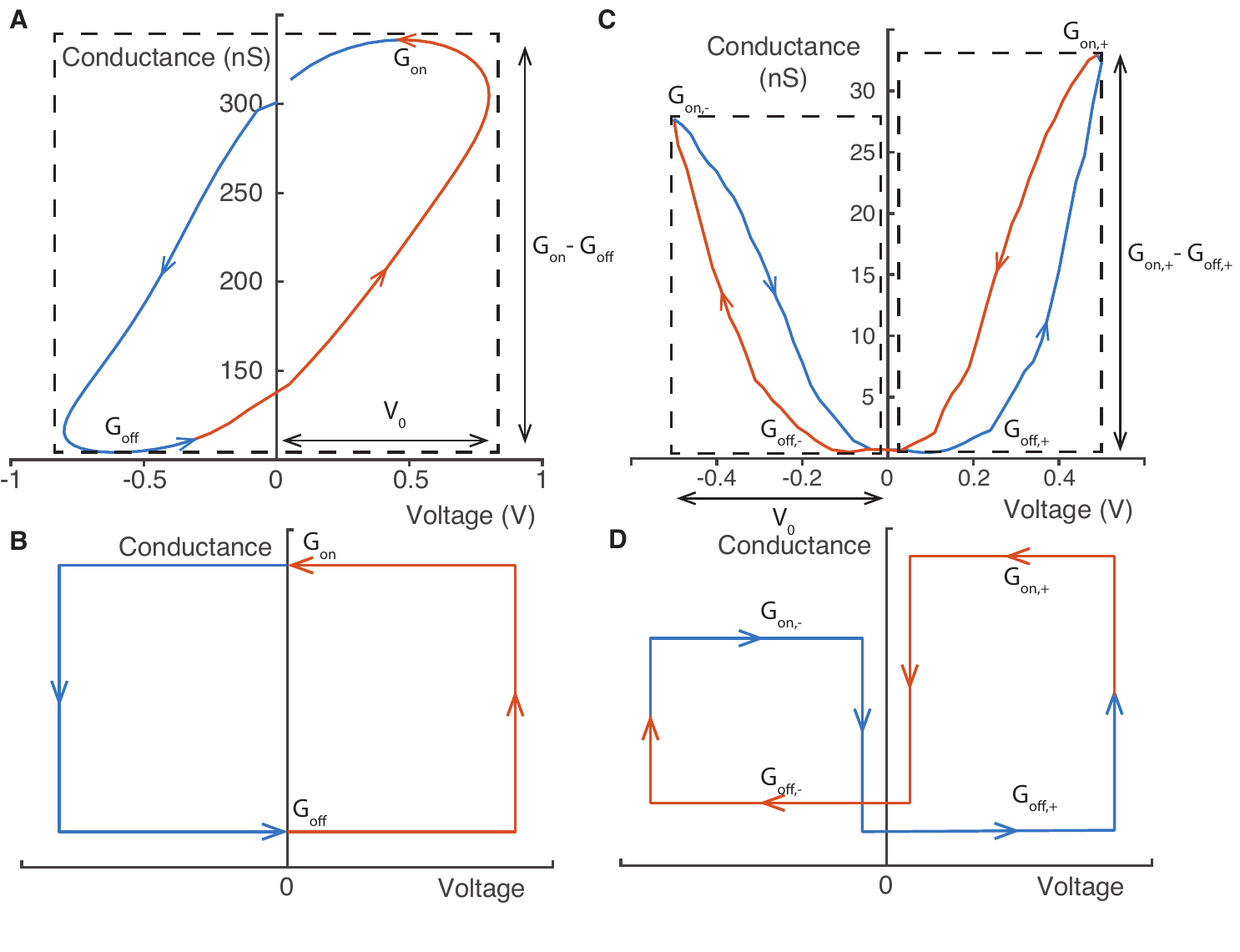}
	\caption*{Fig. S8. \textbf{Normalization of loop area.} (\textbf A) Example of an experimental conductance-voltage curve for a bipolar memristor (see Figure 1C from main text), corresponding to an adsorption-desorption memory mechanism, with the graphical determination of $G_\text{on}$ and $G_\text{off}$.  (\textbf{B}) Idealized loop in the conductance-voltage curve. Its area is used as a normalization factor for a bipolar memristor. (\textbf{C}) Example of an experimental conductance-voltage curve for a unipolar memristor (see Fig. 1B from main text), corresponding to an Wien effect memory mechanism, with the graphical determination of $G_{\text{on},\pm}$ and $G_{\text{off},\pm}$. (\textbf{D}) Idealized loop in the conductance-voltage curve, defining the normalization factor for a unipolar memristor. {Note that the difference between $G_{\text{off},-}$ and $G_{\text{off},+}$ is exaggerated compared to experimental data, to allow easier visualization.}}
\end{figure}

\begin{figure}[h!]
	\centering
	\includegraphics[width=\textwidth]{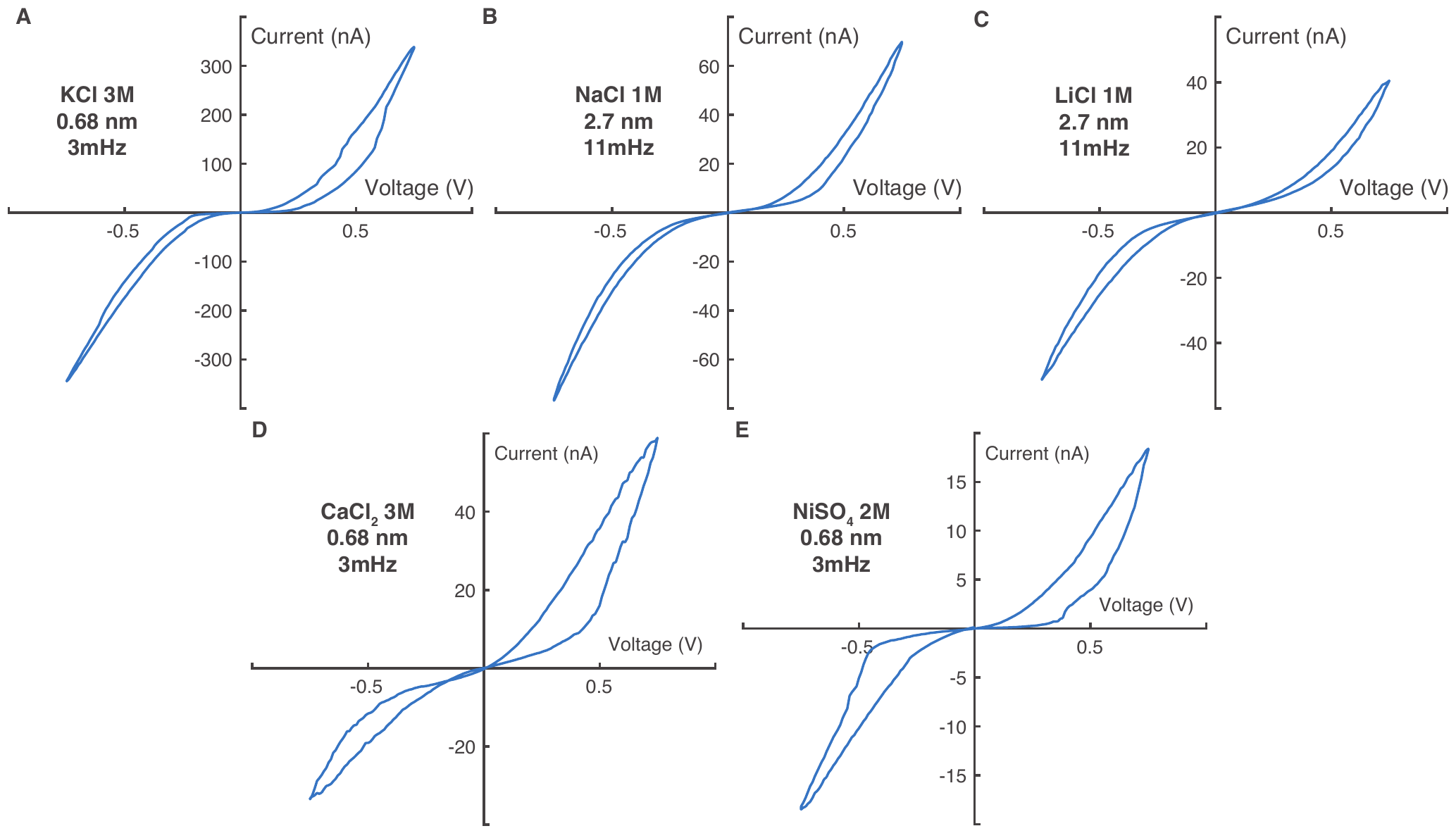}
	\caption*{Fig. S9. \textbf{Evolution of the memristor effect with various electrolytes (pristine channels)}. (\textbf{A} to \textbf{E}) Current-voltage characteristics of an activated carbon channel with height under a voltage sweep of amplitude $0.75 \, \si{V}$. Other parameters (salt type, salt concentration, channel height and frequency) are specified on each panel.}
\end{figure}

\begin{figure}[h!]
	\centering
	\includegraphics[width=\textwidth]{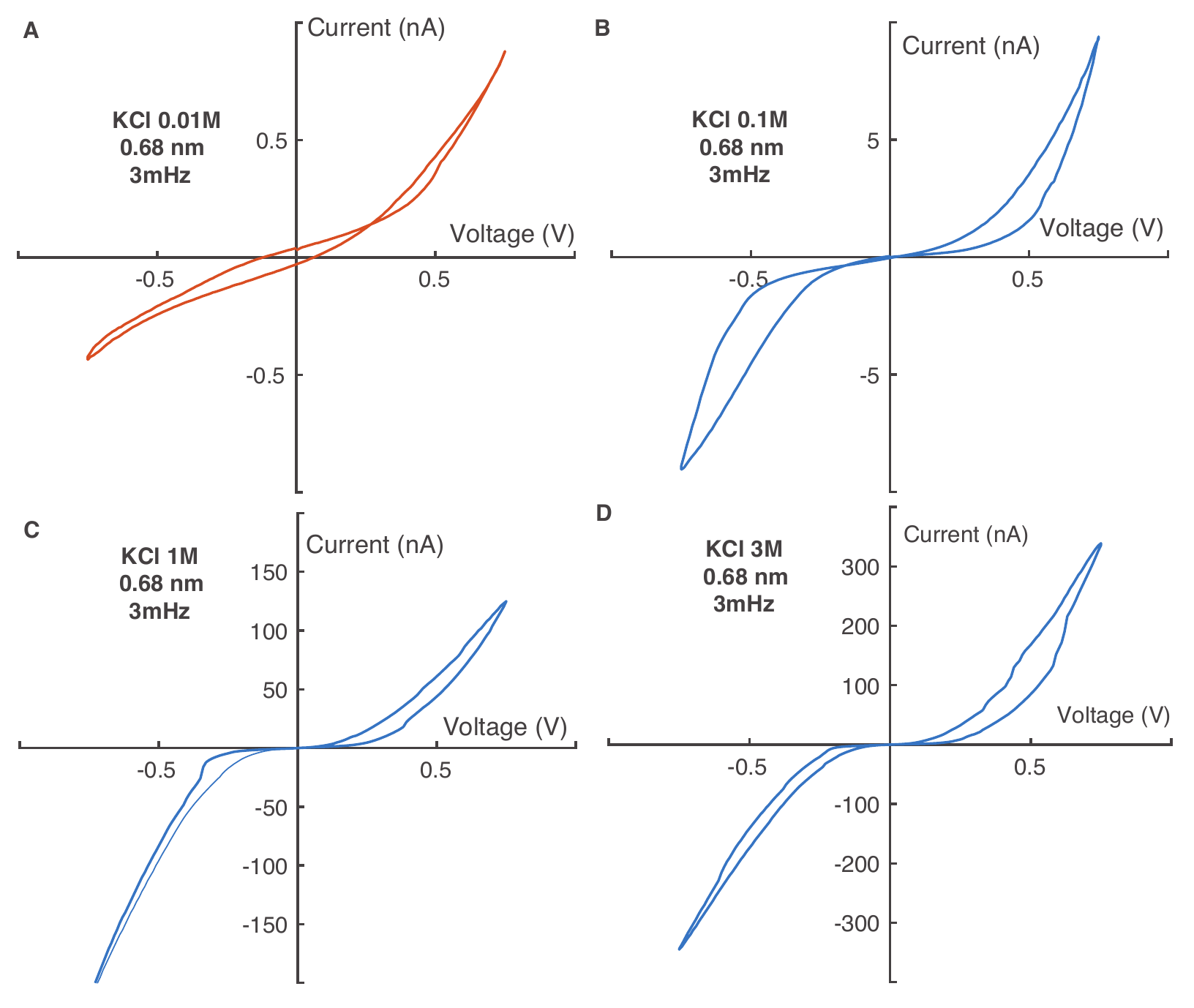}
	\caption*{Fig. S10. \textbf{Evolution of the memristor effect with salt concentration (pristine channels)}. (\textbf{A} to \textbf{D}) Current-voltage characteristics of a pristine MoS$_2$ channel with height $h = 0.68 \, \si{nm}$, filled with potassium chloride at various concentrations, under a voltage sweep of amplitude $0.75 \, \si{V}$ and frequency $3 \, \si{mHz}$. Orange curve indicates a self-crossing loop, while blue curves do not self-intersect.}
\end{figure}

\begin{figure}[h!]
	\centering
	\includegraphics[width=\textwidth]{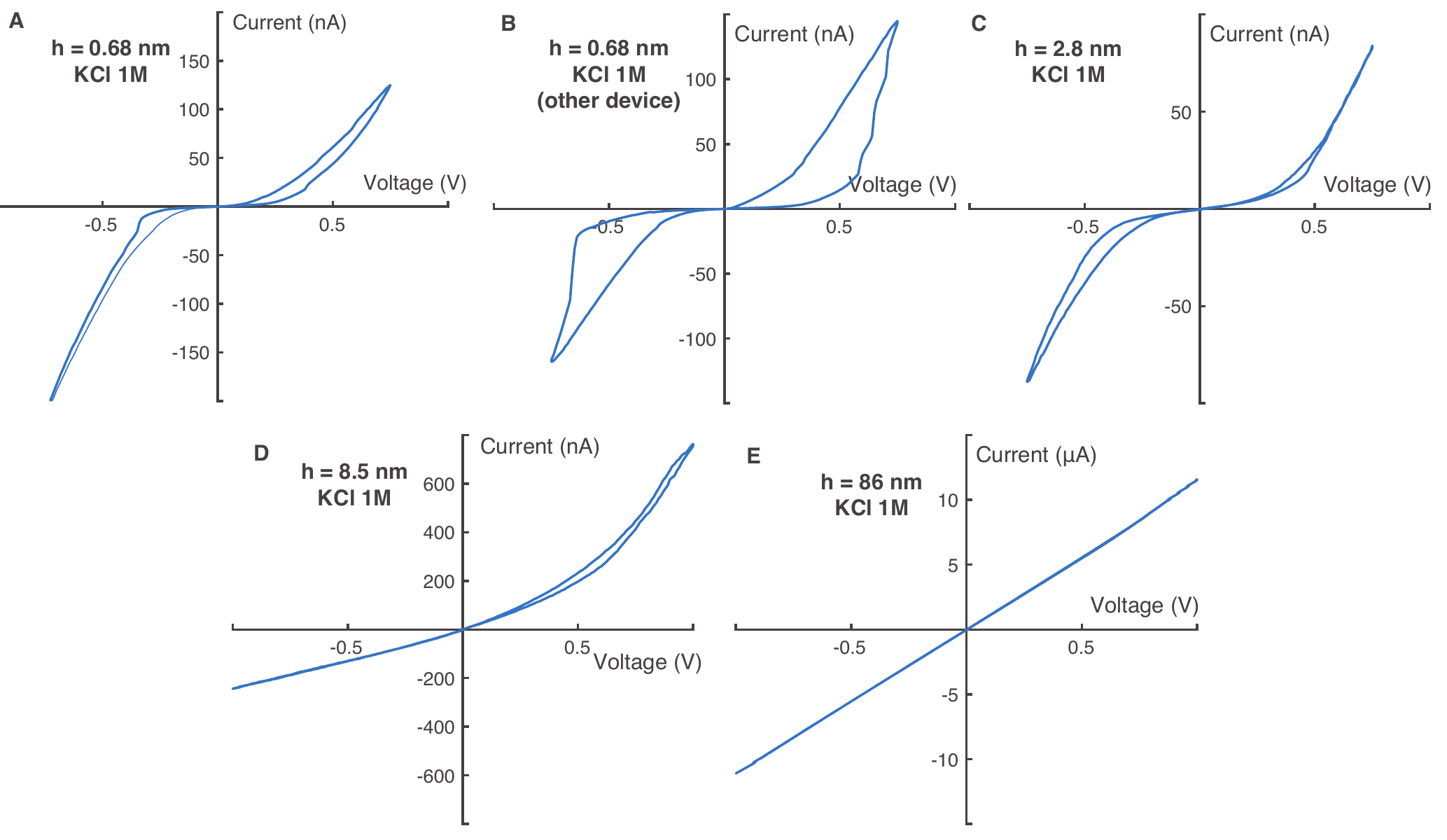}
	\caption*{Fig. S11. \textbf{Evolution of memristive effects with channel height (pristine channels)}. Current-voltage characteristics of pristine MoS$_2$ channels of different heights filled with potassium chloride, under a voltage sweep (frequency $f = 3 \, \si{mHz}$). (\textbf{A} and \textbf{B}) Two different devices with height $h = 0.68 \, \si{nm}$ {(with salt concentration $1 \, \si{M}$, voltage amplitude $0.75 \, \si{V}$)}. (\textbf{C} to \textbf{E}) Devices with height $h = $ 2.8, 8.5 and $86 \, \si{nm}$, respectively ({salt} concentration $1 \, \si{M}$, voltage amplitude $0.75 \, \si{V}$ or $1 \, \si{V}$).}
\end{figure}

\begin{figure}[h!]
	\centering
	\includegraphics[width=\textwidth]{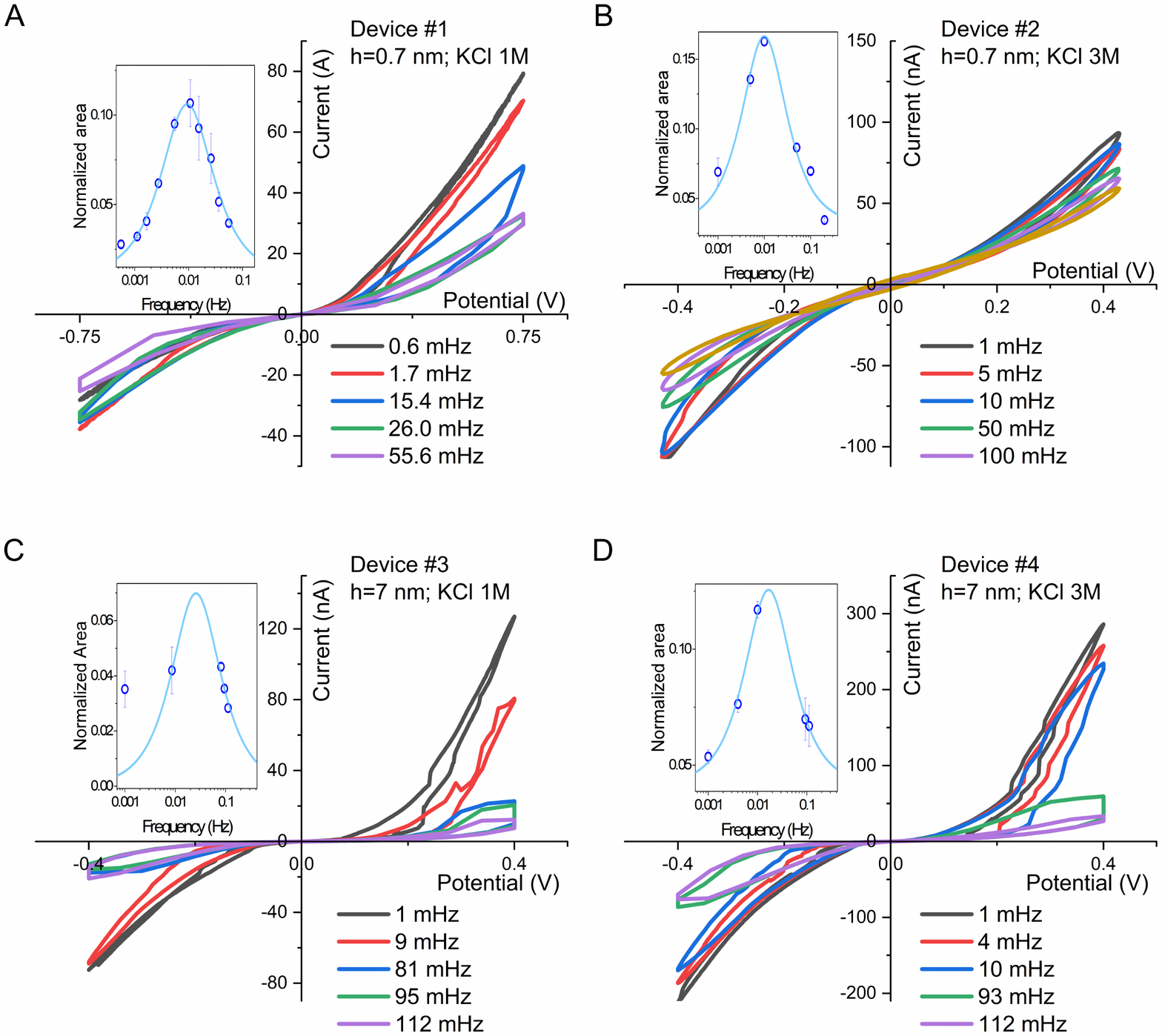}
	\caption*{\label{SI_frequency_activated}Fig. S12. \textbf{Evolution of the memristor effect with voltage frequency (pristine channels)}. Current-voltage characteristics of different pristine MoS$_2$ channels with channel heights $h = 0.68 \, \si{nm}$  (\textbf{A} and \textbf{B}) and $7 \, \si{nm}$  (\textbf{C} and \textbf{D}), the electrolyte is 1M KCl for A, C and 3M KCl for B, D; applied voltage is sinusoidal with frequency ranging from 0.6 mHz to 200 mHz. The normalized loop area vs frequency of \textbf{A} is presented in main Fig. 4C of main text. Insets represent the corresponding normalized areas of the different current-voltage characteristics for that device. The error bars represent the area variation between three successive voltage sweeps.}
\end{figure}

\begin{figure}[h!]
	\centering
	\includegraphics[width=\textwidth]{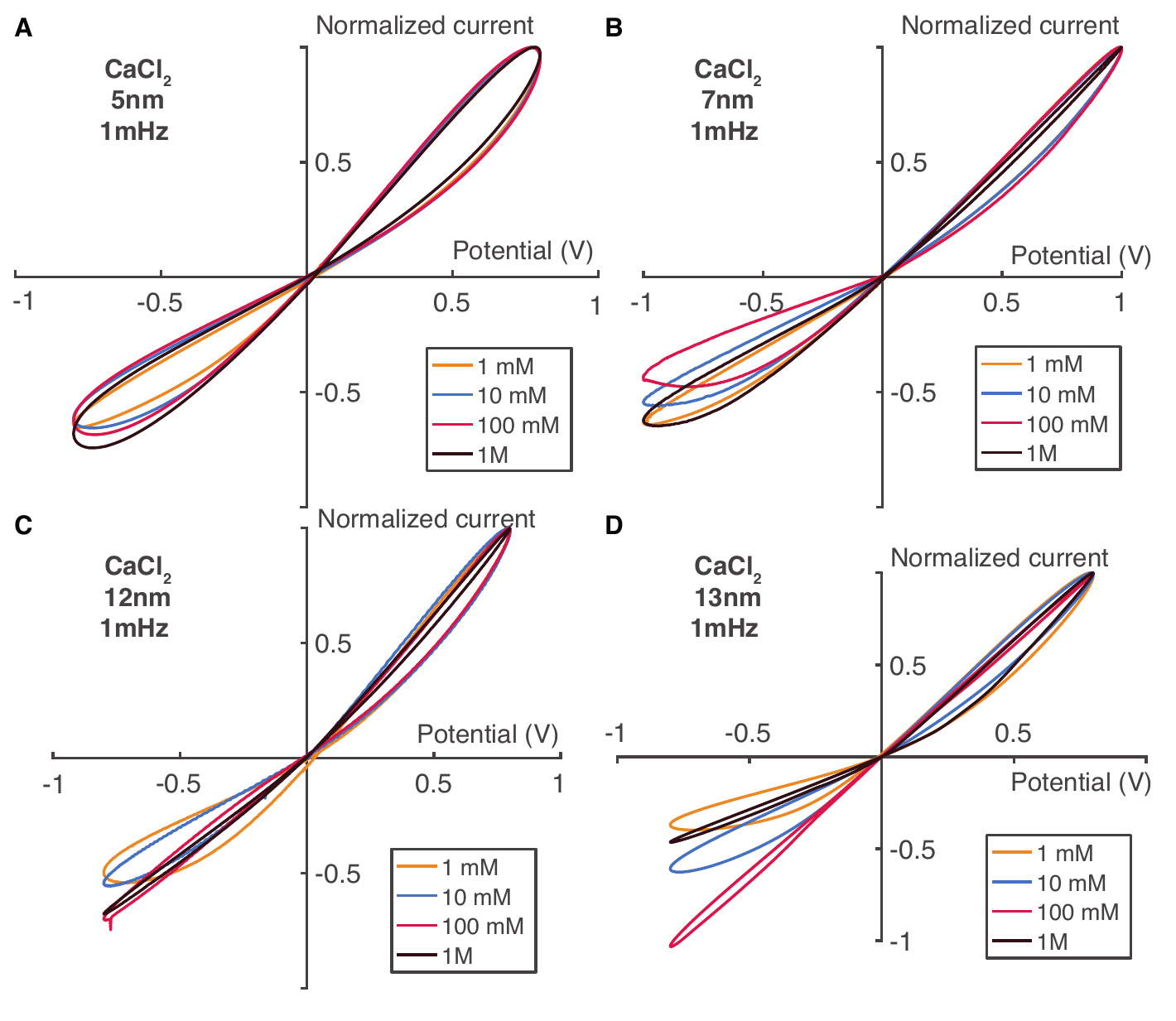}
	\caption*{\label{SI_concentration_activated}Fig. S13. \textbf{Evolution of the memristor effect with salt concentration (activated channels).} (\textbf A to \textbf D) Current-voltage characteristics of four different activated carbon channels, with CaCl$_2$ and AC voltage oscillating between $\pm 0.8 \, \si{V}$ at $1 \, \si{mHz}$. The current is normalized by its maximum value for each salt concentration.  In each case, current is normalized by its maximum absolute value, to allow easier comparison between different datasets.}
\end{figure}

\begin{figure}[h!]
	\centering
	\includegraphics[width=\textwidth]{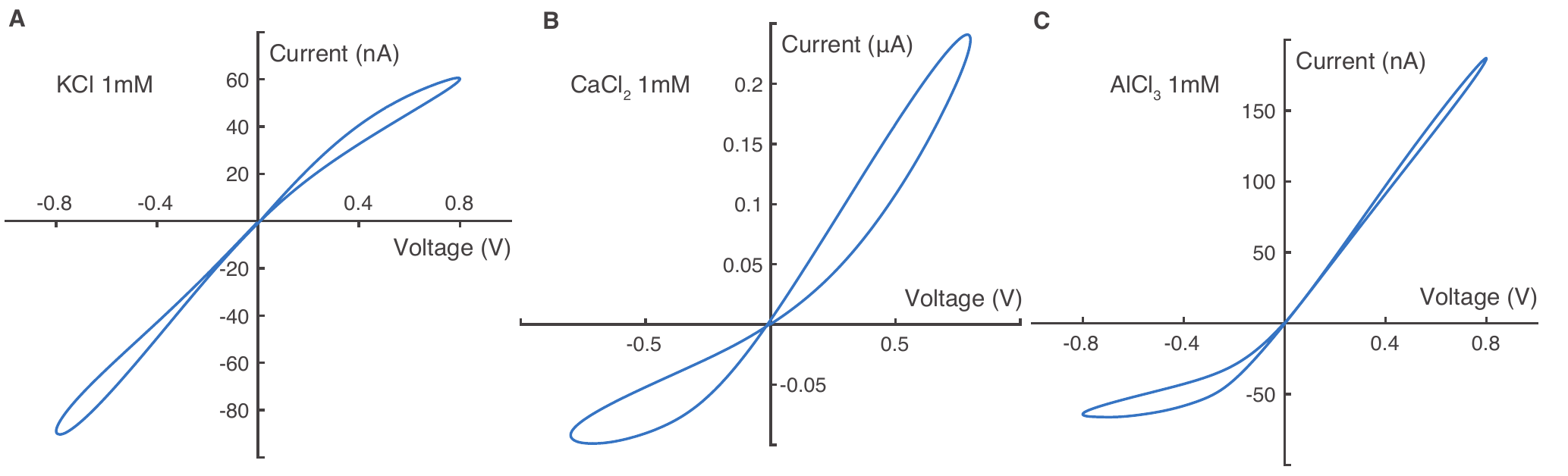}
	\caption*{Fig. S14. \textbf{Evolution of the memristor effect with various electrolytes (activated channels).} (\textbf A to \textbf C) Current-voltage characteristics of an activated carbon channel with height $h = 13 \, \si{nm}$, under a voltage sweep of frequency $1 \, \si{mHz}$ and amplitude $0.8 \, \si{V}$.}
\end{figure}

\begin{figure}[h!]
	\centering
	\includegraphics[width=\textwidth]{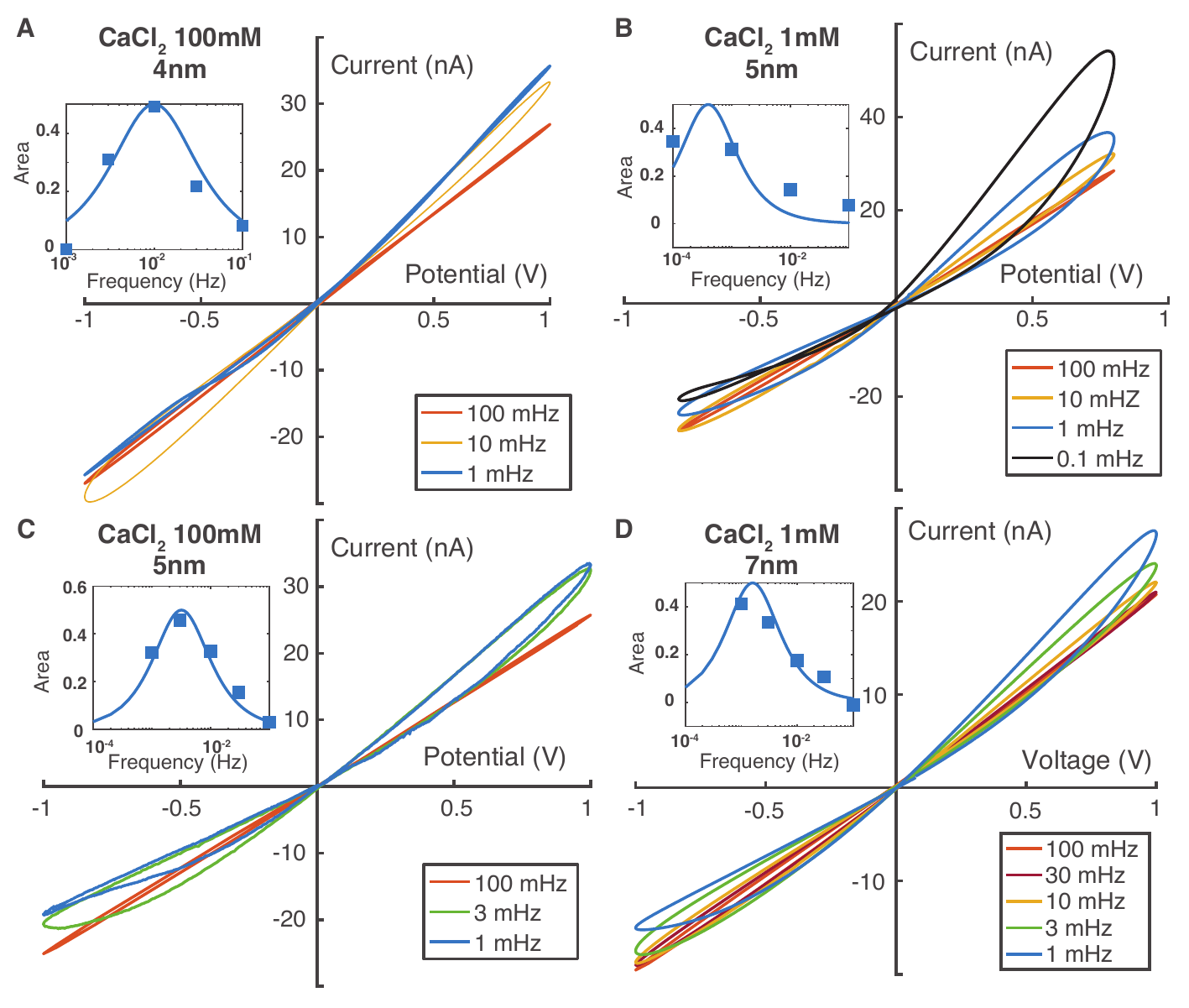}
	\caption*{\label{SI_frequency_pristine}Fig. S15. \textbf{Evolution of the memristor effect with voltage frequency (activated channels).} (\textbf{A} to \textbf{D}) Current-voltage characteristics of four different devices filled with 100 mM or 1 mM CaCl$_2$ (as indicated on each panel). Inset: Normalized area versus frequency. Squares are experimental values and solid lines are theoretical model with memory time parameter, $\tau_m$ equals to $50 \, \si s$ (B),  $100 \, \si s$ (C) and $400 \, \si  s$ (D). The normalized area vs frequency of the $4 \, \si{nm}$ device (A) is presented in main text Fig 4.C.}
\end{figure}

\begin{figure}[h!]
	\centering
	\includegraphics[width=\textwidth]{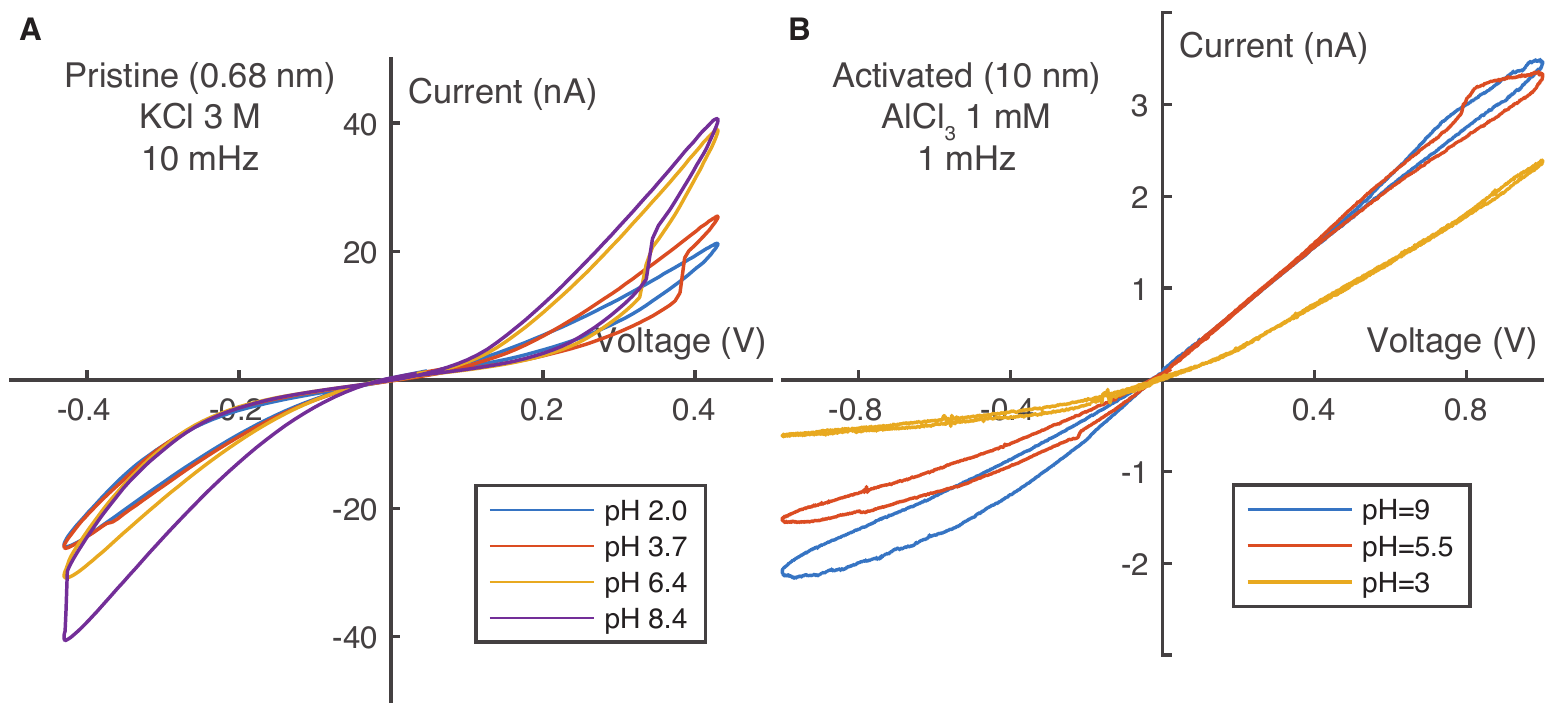}
	\caption*{\label{SI_pH}Fig. S16. \textbf{Evolution of the memristor effect with pH.} (\textbf A) Current-voltage characteristics of a pristine MoS$_2$ channel (height $0.68\,$nm) filled with 3\,M KCl under a voltage sweep at 10\,mHz, for different pH values. (\textbf{B}) IV curve of an activated carbon channel (height $10\,$nm) filled with AlCl$_3$ 1\,mM under a voltage sweep at 1\,mHz, for different pH values.}
\end{figure}

\begin{figure}[h!]
	\centering
	\includegraphics[width=\textwidth]{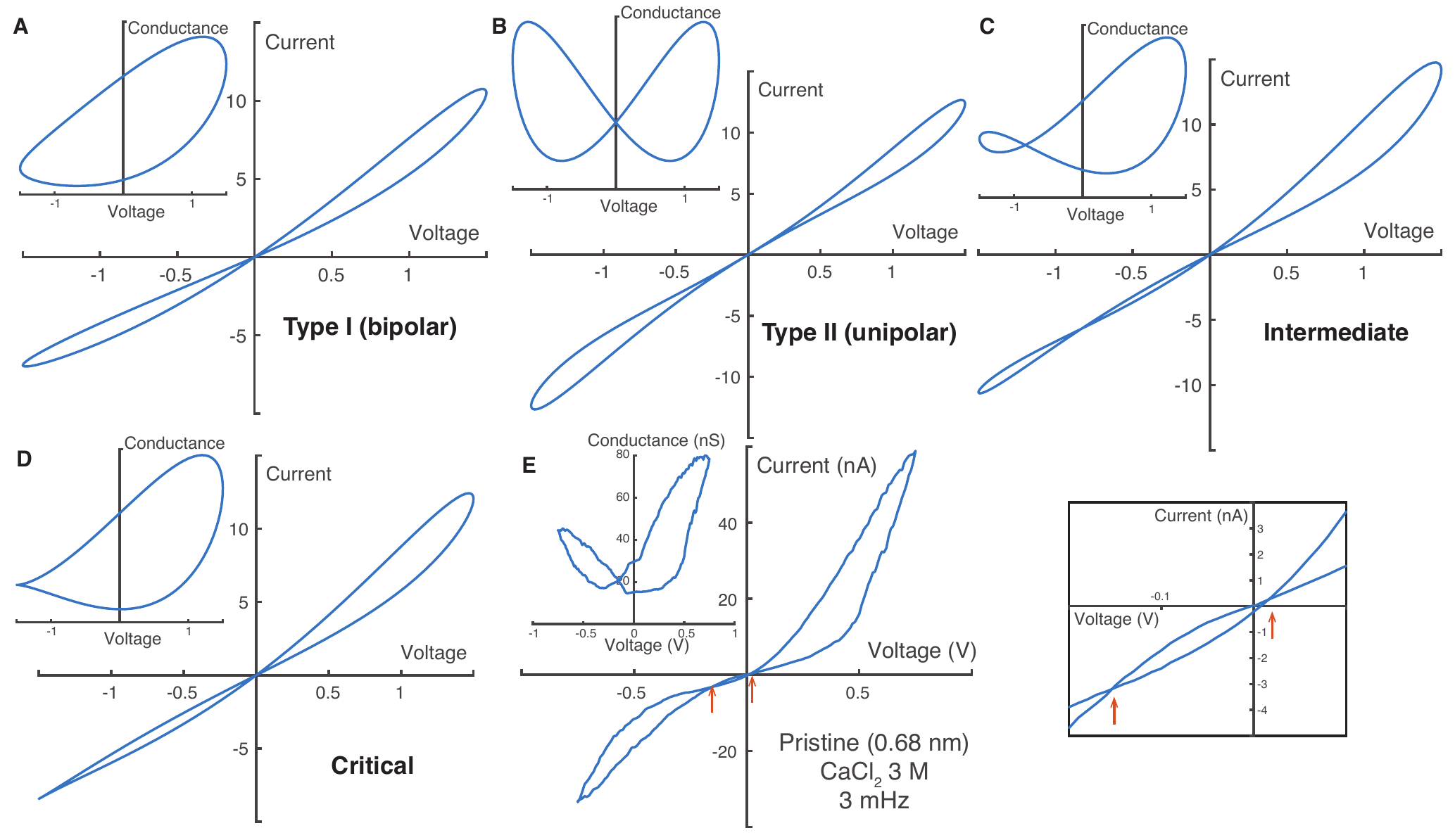}
	\caption*{\label{SI_transition}Fig. S17. \textbf{Transition between bipolar and unipolar behavior.} (\textbf{A} to \textbf{D}) IV and GV (inset) curves for the minimal model of a "mixed" memristor (equations S47-S49). All units are arbitrary. Parameters: $G_\text{unipolar} = 3$, $G_\text{bipolar} = 2$, $\tau =1$, $V(t) = 1.5 \sin t$. (\textbf{A}) $\alpha = 0.8$, $\beta = 0.2$ (bipolar case). The IV curve self-intersects once, but the GV curve does not. (\textbf{B}) $\alpha = 0$, $\beta = 0.8$ (unipolar case). The IV curve does not self-intersect; the GV curve self-intersects on the $y$ axis. (\textbf{C}) $\alpha = \beta = 0.8$ (intermediate case). The IV curve self-intersects twice and the GV curves self-intersects outside of the $y$ axis. (\textbf{D}) $\alpha = 0.8$, $\beta = 0.45$ (criticality). The IV curve self-intersects once and the GV curve develops a cusp. (\textbf{E}) Left panel: Example of experimental IV curve showing a "mixed" memristor behavior, obtained for a pristine MoS$_2$ channel (height $0.68\,$nm) filled with CaCl$_2$ $3\,$M at voltage frequency $3\,$mHz. Red arrows indicate the two crossing points. Inset is the corresponding GV curve. Right panel: Zoom-in on the two crossing points of the IV curve.}
\end{figure}

\clearpage

\begin{DontShowMe}
	\bibliographystyle{./Science}

\end{DontShowMe}